\documentclass[aps,prx,twocolumn,superscriptaddress]{revtex4-2}
\usepackage{graphicx}
\usepackage{bm}
\usepackage{amssymb}
\usepackage{color}
\usepackage{amsmath}
\usepackage{amstext}
\usepackage[english]{babel}
\usepackage{array}             
\usepackage{multirow}         
\usepackage{braket}
\usepackage{physics}
\usepackage[usenames,dvipsnames]{xcolor}
\usepackage[colorlinks=true, citecolor=Blue, linkcolor=RubineRed, urlcolor=Blue]{hyperref}
\usepackage[markup=Highlight, color=yellow, draft]{pdfcomment}
\usepackage{mathrsfs}

\def\zb{{\pmb z}}
\def\xb{{\pmb x}}

\begin{document}
\title{Universal resources for quantum approximate optimization algorithm and quantum annealing}
\author{Pablo D\'iez-Valle\href{https://orcid.org/0000-0001-8338-7973}{\includegraphics[scale=0.45]{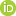}}}
\email{pvalle@itg.es}
\affiliation{Instituto de F\'isica Fundamental IFF-CSIC, Calle Serrano 113b, Madrid 28006, Spain}
\affiliation{Instituto Tecnológico de Galicia, Cantón Grande 9, Planta 3, 15003 A Coruña, Spain}
\author{Fernando J. G\'omez-Ruiz\href{https://orcid.org/0000-0002-1855-0671}{\includegraphics[scale=0.45]{orcid}}}
\email{fegomezr@fis.uc3m.es}
\affiliation{Departamento de F\'isica, Universidad Carlos III de Madrid, Avda. de la Universidad 30, 28911 Legan\'es, Spain}
\author{Diego Porras\href{https://orcid.org/0000-0003-2995-0299}{\includegraphics[scale=0.45]{orcid}}}
\email{diego.porras@csic.es}
\affiliation{Instituto de F\'isica Fundamental IFF-CSIC, Calle Serrano 113b, Madrid 28006, Spain}
\author{Juan Jos\'e Garc\'ia-Ripoll\href{https://orcid.org/0000-0001-8993-4624}{\includegraphics[scale=0.45]{orcid}}}
\email{juanjose.garcia.ripoll@csic.es}
\affiliation{Instituto de F\'isica Fundamental IFF-CSIC, Calle Serrano 113b, Madrid 28006, Spain}
\begin{abstract}
The Quantum Approximate Optimization Algorithm (QAOA) is a variational ansatz that resembles the Trotterized dynamics of a Quantum Annealing (QA) protocol. This work formalizes this connection formally and empirically, showing the angles of a multilayer QAOA circuit converge to universal QA trajectories. Furthermore, the errors in both QAOA circuits and QA paths act as thermal excitations in pseudo-Boltzmann probability distributions whose temperature decreases with the invested resource---i.e. integrated angles or total time---and which in QAOA also contain a higher temperature arising from the Trotterization. This also means QAOA and QA are cooling protocols and simulators of partition functions whose target temperature can be tuned by rescaling the universal trajectory. The average cooling power of both methods exhibits favorable algebraic scalings with respect to the target temperature and problem size, whereby in QAOA the coldest temperature is inversely proportional to the number of layers, $T\sim 1/p$, and to the integrated angles---or integrated interactions in QA.\\
\\
Ref.:~\href{https://doi.org/10.1103/hxv2-sbr7}{Phys. Rev. Research {\bf 8}, 013211 (2026)}
\end{abstract}
\maketitle

\section{Introduction}
In quantum computing and quantum simulation, one of the most interesting and difficult tasks is the preparation of ground states of interacting Hamiltonians. The Hamiltonian $\hat{H}_{\rm QSNet}$ of the Ising quantum spin network (QSNet) can be mapped onto classical optimization problems of the Quadratic Unconstrained Binary Optimization (QUBO) or higher-order unconstrained binary optimization type, demonstrating that the task is at least NP-hard. Two methods with very similar inspiration have been proposed to solve such problems: Quantum Annealing (QA)~\cite{Aharonov04,Lidar_RMP18} and the Quantum Approximate Optimization Algorithm (QAOA)~\cite{Farhi14A,Farhi14B,Farhi_Q22}. QA is a continuous protocol based on a slow deformation of the system's dynamics, from a {\it mixing Hamiltonian}, typically denoted as $\hat{H}_x$, whose ground state is initially prepared, to the {\it cost Hamiltonian}, $\hat{H}_\text{QSNet}$,  whose ground state we wish to achieve~\cite{Lidar_RMP18}. QAOA mimics a Trotterized version of QA, engineering a quantum circuit with $p$ layers that alternate evolution with the mixing and cost Hamiltonians. The rotation angles of this variational circuit are then optimized to best approximate the ground state~\cite{Farhi14A,Farhi14B,Farhi_Q22}.

The performance and scaling of resources in QA and QAOA protocols remains an open question. When focusing on QA, studies typically have sought guarantees of success using the adiabatic theorem~\cite{Aharonov04}. However, pursuing this sufficient condition and optimizing the adiabatic trajectories require a very good knowledge of the spectral gaps of many-body systems~\cite{Susa_PRA21,Helmut_PRR24, Chen_NatMach2022}, which is harder than solving the original problem. QAOA on the other hand is already based on the design of optimal trajectories and addressing the same question seems more accessible, thanks to a reduced dimensionality of the control space and access to emulators, computers and simulators of increasing size~\cite{Pelofske_2023,Pelofske2024,he_2024,Monroe_PNAS20}. However, despite the relative success in heuristic strategies---warm-start initializations of parameters~\cite{Egger_warm_2021, tate2024theoreticalapproximationratioswarmstarted}, problem symmetries~\cite{lyngfelt2025symmetry}, extrapolation~\cite{Zhou_PRX20, Lee_depth_2023, Shaydulin_parameter_2023}, or machine learning techniques~\cite{Deshpande_capturing_2022,cheng_quantum_2024,giovanni_genetic_2023}---to overcome the NP-hard nature of variational quantum optimization~\cite{Bittel_training_2021} 
and the problems of local minima and barren plateaus, there are no conclusive results yet about the performance and asymptotic trends of this algorithm, either.

In this work we adopt a different strategy, which combines an empirical understanding of the states created by quantum optimization, with a study of emergent universal properties common to QAOA and QA. The first central result in our work is showing that the concentration of parameters in optimal QAOA circuits~\cite{Farhi_2022, Basso_quantum_2022, Lee_depth_2023, Zhou_PRX20, Akshay_parameter_2021, Claes_2021} happens on universal QA passages of monotonically growing interaction and progressively decreasing mixer Hamiltonian. We show that these multilayer QAOA and QA protocols with said trajectories act as cooling protocols that create pseudo-Boltzmann probability distributions - i.e., distributions that fluctuate around the Boltzmann distribution in energy space. This result goes beyond our previous work with single-layer circuits~\cite{DiezValle_PRL23, DiezValle_Frontiers2024}, providing evidence that the non-adiabatic errors in QAOA and QA passages behave as thermal contributions, with the peculiarity that QAOA has an additional but vanishing error that originates in the Trotterized evolution. Our discovery enables the application of QAOA and QA protocols without previous optimization and their use as a cooling algorithm or simulators of partition functions with tuneable temperature. Finally, this study is supported by an interpretation of both multilayer QAOA and QA as paths in Hamiltonian space, deriving common metrics for the resources associated to these two protocols. We provide evidence that these resources---integrated angles, time, integrated interactions, and number of layers---scale very favorably with problem size and target state temperature.

The property of angle concentration in QAOA optimized trajectories is not a new discovery. This result has been demonstrated for MaxCut on regular graphs and the Sherrington-Kirkpatrick model~\cite{Farhi_2022}, and remarkable patterns have been repeatedly reported in multiple studies~\cite{Basso_quantum_2022, Lee_depth_2023, Zhou_PRX20, Akshay_parameter_2021, Claes_2021}. Furthermore, the transfer of angles between problems has also been used in the past among problems of similar families~\cite{Galsa_transferability_2021, Shaydulin_parameter_2023, shaydulin2024, Sureshbabu_2024}, and there is evidence of smooth annealing paths constructed from QAOA optimal parameters~\cite{Zhou_PRX20}. However, our work is the first one to demonstrate the universality of these concentrated paths - universality defined over the ensembles of dense spin glasses - and their connection to continuous QA trajectories, while at the same time identifying the errors of those discretized annealing protocols as quasithermal distributions with tuneable properties, extending our physical interpretation of QAOA~\cite{DiezValle_PRL23, DiezValle_Frontiers2024} to both QAOA and QA simultaneously.

The paper is organized as follows. In Sec.~\ref{sec_QAOA}, we refresh the main elements of a QAOA circuit and the Hamiltonians over which it is defined. Then, in Sec.~\ref{sec:qaoa-with-qa} we introduce a formal connection between QAOA evolution and QA passages, introducing a notion of resource or cost that is common to both protocols. Sec.~\ref{sec_qaoacooling} presents a physical analysis of the states created by multilayer QAOA, demonstrating that this circuit acts as a cooling protocol that creates bimodal Boltzmann distributions in energy space. This study conveniently provides us with two metrics of ``quality'' of QAOA, in the form of the two temperatures of the distribution---the cold temperature that determines our probability to approximate the ground state, and the hotter temperature for the background noise. Sec.~\ref{sec_universalAQCtrajectories} builds on previous results, computing optimal QAOA with different number of layers, over hundreds of different problems with up to $N=20$ qubits. This study provides evidence that QAOA optimal parameters converge to universal trajectories that resemble continuous QA passages.  Sec.~\ref{sec_scalingresources} analyzes these circuits, studying the resources demanded by the QAOA circuits to achieve distributions with progressively lower temperatures, and showing very promising trends. We conclude with a summary of the main results and a discussion of future research in Sec.~\ref{sec_conclusions}.

\section{Quantum Approximate Optimization Algorithm}
\label{sec_QAOA}

\begin{figure}[t!]
\includegraphics[width=1.0\linewidth]{ 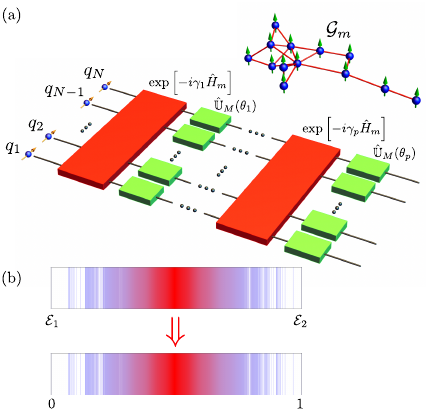}
\caption{\label{fig_1} (Color online) {\bf Overview of the QAOA algorithm with $p$-layers, depicted schematically.} In panel (a), we present a $p$-layer QAOA algorithm. The red and green modules represent the cost and mixer layers, respectively. The cost layer is defined by a quantum spin network Ising Hamiltonian, as described in Eq.~\eqref{H_QSNet}. Each experimental sample, denoted as the $m$-th sample, corresponds to a quantum spin network with topology $\mathcal{G}_m$. Panel (b) illustrates the rescaling of the energies in Eq.~\eqref{eq_energyrescaling}, which maintains the same energy levels structure.}
\end{figure}

QAOA is a hybrid quantum-classical algorithm used for solving combinatorial optimization problems~\cite{Farhi14A,Farhi14B}. The quantum part of this algorithm can be implemented on a programmable gate-based universal quantum computer. The QAOA algorithm involves preparing the initial state $\ket{\psi_+}$, followed by applying the single-layer QAOA operator for $p$ rounds, given by:
\begin{equation}\label{eq_QAOAstate}
\ket{\pmb \gamma_p,\pmb \theta_p}=\prod_{i=1}^{p}\hat{\mathcal{Q}}_i (\gamma_i,\theta_i)\ket{\psi_+}.
\end{equation}
Initially, all qubits are prepared in a superposition state encompassing all possible solutions, as given by the application of $N$ Hadamard gates $\mathbb{H}$ onto the qubits' ground states $\ket{0}$
\begin{equation}
\ket{\psi_{+}}:=\ket{+}^{\otimes N}=\mathbb{H}^{\otimes N}\ket{0}^{\otimes N}
\end{equation}
For simplicity of notation and ease of understanding, we can gather all the $\gamma_i$ and $\theta_i$ values in $p$-dimensional vectors denoted by $\pmb \gamma_{p} = (\gamma_1,\ldots,\gamma_p)$ and $\pmb \theta_{p} = (\theta_1,\ldots,\theta_p)$. This state is evolved under the circuit defined by Eq.~\eqref{eq_QAOAstate} with variational parameters $[\pmb \gamma_p, \pmb \theta_p]$ that are tuned to minimize the expectation value of the energy. Thus, we have:
\begin{equation}\label{VP_QAOA}
\left\{\pmb \gamma_p, \pmb \theta_p\right\}=\arg \min \left[\bra{\pmb \gamma_p,\pmb \theta_p} \hat{H}_{\rm QSNet} \ket{\pmb \gamma_p,\pmb \theta_p}\right].
\end{equation}
The set of variational parameters $\{\pmb \gamma_p, \pmb \theta_p\}$ is estimated through a classical optimization process, and measuring the state $\ket{\pmb\gamma_p, \pmb\theta_p}$ on the computational basis provides an approximate solution to the optimization problem encoded by $\hat{H}_{{\rm QSNet}}$.

 The operator $\hat{\mathcal{Q}}_i$ corresponds to the $i$-th layer of the QAOA. The single-layer QAOA operator is defined as:
\begin{equation}\label{QAOA_Ly_Oper}
\hat{\mathcal{Q}}_i (\gamma_i,\theta_i)=\hat{\mathbb{U}}_{M}(\theta_i)\exp\left[{-i\gamma_i \hat{H}_{\rm QSNet}}\right],
\end{equation}
where $\hat{\mathbb{U}}_{M}(\theta_i)$ is referred to as the mixer layer. Traditionally, the mixer layer is generated by an $x$ rotation on every qubit, defined as:
\begin{equation}
    \hat{\mathbb{U}}_{M}(\theta_i)
    = \exp\left[-i\frac{\theta_i}{2} \hat{H}_x\right],
\end{equation}
where $\hat{H}_x\equiv \sum_{n=1}^N \hat{\mathbb{X}}_{n}$ is the mixing Hamiltonian. The second term in Eq.~\eqref{QAOA_Ly_Oper} is called the cost layer. The exact form of the cost layer is problem-dependent and characterized by the quantum spin network Ising Hamiltonian underlying the combinatorial optimization problem. A quantum spin network (QSNet) is defined by an undirected graph $\mathcal{G}(\mathcal{V},\mathcal{E})$, where the number of vertices is equal to the number of qubits or spins $|\mathcal{V}|=N$, and the edge weight for the vertices $(n,m)\in(\mathcal{E})$ is $J_{nm}$. The cost Hamiltonian can be written as:
 \begin{equation}\label{H_QSNet}
\hat{H}_{{\rm QSNet}}=\sum_{(n,m)\,\in\, \mathcal{E}} J_{nm} \hat{\mathbb{Z}}_{n}\hat{\mathbb{Z}}_{m}+\sum_{n\,\in\,\mathcal{V}}h_n  \hat{\mathbb{Z}}_{n},
\end{equation}
where $\pmb J$ and $\pmb h$ are an $N$-by-$N$ square coupling matrix and a vector of $N$ coefficients respectively, with $N$ the number of qubits or spins, and $J_{nm}$, $h_n$ are the matrix and vector coefficients. The operators $\hat{\mathbb{X}}_{i}$ and $\hat{\mathbb{Z}}_{i}$ are matrices of order $2^N$ defined by the relations $\hat{\mathbb{X}}_{i}=\hat{\mathbb{I}}_{1}\otimes\ldots\otimes\hat{\mathbb{I}}_{i-1}\otimes\hat{\sigma}_{i}^{x}\otimes\hat{\mathbb{I}}_{i+1}\otimes\ldots\hat{\mathbb{I}}_{N}$ and $\hat{\mathbb{Z}}_{i}=\hat{\mathbb{I}}_{1}\otimes\ldots\otimes\hat{\mathbb{I}}_{i-1}\otimes\hat{\sigma}_{i}^{z}\otimes\hat{\mathbb{I}}_{i+1}\otimes\ldots\hat{\mathbb{I}}_{N}$. Here, $\hat{\sigma}_{i}^{\alpha}$ denotes the Pauli operator at site $i$-th along the direction $\alpha=x,z$, $\hat{\mathbb{I}}_i$ is the identity matrix of order 2 at site $i$. In Fig~\ref{fig_1}(a), we schematically represent the QAOA algorithm and a QSNet featuring a graph $\mathcal{G}_m$, where $m$ enumerates the number of numerical experimental implementations.

Many interesting real-world optimization problems are contained in families of NP-hard binary problems such as QUBO~\cite{kochenberger2014}. QUBO problems have an associated classical energy defined by:
\begin{equation}
E_{{\rm QUBO}}(\xb) = 2\sum_{n,m=1}^{N}x_{n}Q_{nm}x_{m},
\label{EQubo}
\end{equation}
where $\mathbf{\xb} = (x_1,\ldots,x_N)$ with $x_i \in \{0,1\}$, and $\pmb Q$ is an $N$-by-$N$ square symmetric matrix of real coefficients denoted by $Q_{nm}$. It is well known that these problems can be mapped to the Ising form~\eqref{H_QSNet}. The binary variables are cast into $\hat{\mathbb{Z}}_{n}$, where $\hat{\mathbb{Z}}_{n}\ket{z_n}=z_n \ket{z_n}$, and $z_n = 2 x_n -1 \in \{-1,+1\}$ denotes the spin value. The map produces a relationship between the variables of Eq.\eqref{EQubo} with the QSNet Hamiltonian parameters given by $J_{n\neq m}=Q_{nm}$, $J_{nn} =0$, and $h_n = \sum_{m=1}^{N}Q_{nm}$. Our study is based on an extensive statistics of randomly generated QUBO problems defined on fully connected graphs, i.e. all off-diagonal elements of $\mathbf{Q}$ or $\mathbf{J}$ are nonzero. The nonzero values $Q_{nm}$ of each sample are randomly drawn from a normal distribution $\mathcal{N}(\mu,\sigma^2)$ with mean $\mu = 0$ and variance $\sigma^2 =1$. We are interested in studying the main characteristics of individual experiments and the collective mean behavior across $m$ samples of QSNets.

\section{Connection between QAOA and Quantum Annealing}
\label{sec:qaoa-with-qa}
The sequence of gates in the protocol~\eqref{eq_QAOAstate} resembles a first order Trotter approximation of a quantum annealing protocol. Quantum annealing is usually formulated as an adiabatic passage with a time-dependent Hamiltonian that progressively deemphasizes a mixing term $\hat{H}_x$, while simultaneously activating the problem Hamiltonian $\hat{H}_{{\rm QSNet}}$ whose ground state we wish to construct. This dynamics can be formulated in terms of a trajectory in Hamiltonian space,
\begin{equation}
  \hat{H}(s) = -A(s) \hat{H}_x + B(s) \hat{H}_{{\rm QSNet}},
  \label{eq:annealinghamiltonian}
\end{equation}
parameterized by a dimensionless quantity $s\equiv t/t_a \in \left[0,1\right]$, that is the time rescaled with respect to the total duration of the protocol $t_a$. Typically, $A(s)$ is a monotonously decreasing and $B(s)$ a monotonously increasing function, which we may set to start at $A(0)=1$, $B(0)=0$ and end at $A(1)=0$, $B(1)=1$. The quantum annealing protocol starts with an easy to prepare ground state of the mixer Hamiltonian, modifying the Hamiltonian over a long time to approximate the ground state of $\hat{H}_\text{QSNet}$ and thus obtain the solution to a hard computational problem~\cite{Lidar_RMP18,Lidar_PRL08,Lidar_PRL02}. In the Schrödinger picture this corresponds to
\begin{equation}
  i\partial_t \ket{\psi}=\hat{H}(t/t_a)\ket{\psi},\quad \ket{\psi(0)} = \ket{\psi_+}.
\end{equation}
As a remark, let us note that we can rescale the time, $t = \alpha \tau$,
\begin{equation}
  i\partial_\tau \ket{\psi}=\alpha \hat{H}(\tau/t_a^{\prime})\ket{\psi},
\end{equation}
reducing the duration of the protocol $t_{a}^{\prime}=t_a/\alpha$ at the cost of enhancing the strength of the Hamiltonians $\alpha\times\hat{H}$. This means that both $t_a$ and $\Vert{\hat{H}}\Vert$ act as interdependent measures of the resources used in the QA protocol.

To avoid this ambiguity and connect the QA dynamics to the QAOA protocol, we may introduce a formulation that does not depend on time, but on the integrated weights of the mixer and interaction Hamiltonian contributions. If we introduce the new independent variable
\begin{equation}
  \Theta(t) = \int_0^t A(\tau)\mathrm{d}\tau,
\end{equation}
then the Schrödinger equation becomes
\begin{equation}
  i \frac{d\Theta}{dt} \partial_\Theta \ket{\psi(\Theta)} = \left[B(\Theta)\hat{H}_{{\rm QSNet}} - \frac{d\Theta}{dt}\hat{H}_x \right]\ket{\psi(\Theta)}.
\end{equation}
Let us define
\begin{equation}
  \Gamma(\Theta) = \int_0^\Theta B(\tau(\Theta_1)) \left(\frac{d\Theta}{d\tau}\right)^{-1}\mathrm{d}\Theta_1,
\end{equation}
so that the Schrödinger equation is
\begin{equation}
  i \partial_\Theta \ket{\psi(\Theta)} = \left[ \frac{d\Gamma}{d\Theta} \hat{H}_{{\rm QSNet}} - \hat{H}_x \right]\ket{\psi(\Theta)}.
  \label{eq_aqcshro}
\end{equation}
Now the QA passage is fully determined by the trajectory $(\Theta, \Gamma)$ in Hamiltonian space, in a way that is totally symmetric---i.e., we would arrive at the complementary equation if we took $\Gamma$ as independent variable. If we fix $\Vert{\hat{H}_x}\Vert$ and $\Vert{\hat{H}_\text{QSNet}}\Vert$, this allows us to define the final values of those two coordinates $(\Theta_\text{max}, \Gamma_\text{max})$ as the resources of the annealing protocol.

This formulation allows establishing an immediate connection to the QAOA algorithm. The QAOA circuit consists of a product of unitaries
\begin{equation}
  \ket{\psi(t_a)} := \prod_{n=1}^{p} e^{-i \frac{1}{2} \theta_n \hat{H}_x}e^{-i \gamma_n \hat{H}_{{\rm QSNet}}}\ket{\psi(0)}
\end{equation}
that implement a first order Trotter approximation of the evolution with the composite Hamiltonian $\hat{H}(\Theta)$. The QAOA angles are the discrete counterparts of the QA Hamiltonian path, with the identification
\begin{align}
  \Theta_n &:= \sum_{m<n} \frac{1}{2} \theta_m\simeq \frac{1}{2} \Theta(\tau_n),\label{eq_Thetan}\\
  \Gamma_n &:= \sum_{m<n}\gamma_n \simeq \Gamma(\tau_n).\label{eq_Gamman}
\end{align}
This allows us to define the resource of a QAOA protocol similarly as the integrated angles $\Theta_{\textnormal{max}} = \Theta_{p}$ and $\Gamma_{\textnormal{max}} = \Gamma_{p}$ over the $p$ circuit layers. We will use this identification of resources in later sections, when we analyze the equivalent of a ``time-to-solution'', connecting the QAOA resources to the quality of the ground state approximation or the effective temperature of the produced states.
\begin{figure*}[t]
\includegraphics[width=0.9\linewidth]{ 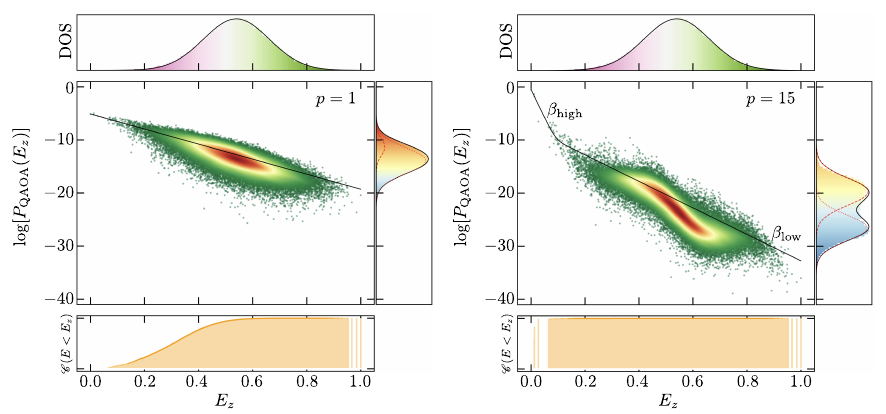}
\caption{\label{fig_QAOA_distribution}(Color online) Optimized QAOA state for one problem with $N=16$ qubits and a circuit with $p=1$ (left panel) and $p=15$ layers  (right panel). We plot the numerical probability amplitude of the QAOA state as dots, together with a solid line representing the best fit to a pseudo-Boltzmann distribution~\eqref{eq_bimodalBoltzmann}, both as a function of the normalized energy $E_z$. The upper histograms show the density of states. The right-side histograms depict the contributions to the probability of the hot and cold components, with a Gaussian fits revealing signatures of the bimodal distribution. The bottom panels show the cumulative probability $\mathscr{C}(E < E_z)$, to illustrate the dominant contribution of the $\beta_\text{high}$ component.}
\end{figure*}

\section{Multilayer QAOA studies}
This section discusses the performance of optimized QAOA circuits over multiple random QUBO problems, sampled as per the description in Sec.~\ref{sec_QAOA}, with increasing number of qubits from $N=2$ to $20$, and up to $p=30$ layers of gates. This extensive study, has been performed over hundreds of problems---800 random problems for $N<20$ and 500 for $N=20$---, obtaining very good statistics for both the type of quantum states produced (c.f. Sec.~\ref{sec_qaoacooling}) as well as for the angles that result from the optimization (c.f. Sec.~\ref{sec_universalAQCtrajectories}).

\subsection{QAOA is a cooling protocol}
\label{sec_qaoacooling}
Inspired by earlier works~\cite{DiezValle_PRL23, DiezValle_Frontiers2024, Lotshaw_PRA23}, we will analyze the output of optimal QAOA multilayer circuits by studying the probability distribution $P(E)$ of the output states in energy space. This is given by
\begin{equation}
  P(E):=\sum_{\zb}\delta (E-E_\zb)P_{{\rm QAOA}}(E_\zb),
  \label{eq_probpE}
\end{equation}
with the probability amplitude
\begin{equation}
P_{{\rm QAOA}}(E_\zb):= \vert\langle \zb\, \lvert {\pmb \gamma}_p, {\pmb \theta}_p\rangle\vert^2
\end{equation}
derived from the $p$-layer QAOA state~\eqref{VP_QAOA}. For ease of presentation and to allow comparison among different problem Hamiltonians, we will typically rescale the energies of the different states $E_\mathbf{z}$, so that $E_\mathbf{z}=0$ corresponds to the ground state and $E_\mathbf{z}=1$ corresponds to the highest eigenvalue of $\hat{H}_\text{QSNet}$. That is
\begin{equation}
  E_\mathbf{z} = \frac{E_\mathbf{z}-E_\text{min}}{E_\text{max}-E_\text{min}},
\label{eq_energyrescaling}
\end{equation}
where $E_\text{max}$ and $E_\text{min}$ are the largest and smallest eigenvalue of the specific problem $\hat{H}_\text{QSNet}$ under study (see Fig.~\ref{fig_1}b).

Fig.~\ref{fig_QAOA_distribution} shows two typical examples of the probability distribution $P_\text{QAOA}$ associated to QAOA circuits optimal angles, for $p=1$ and $p=15$ layers. As expected, larger numbers of layers imply both larger values of the resources $\Theta_{\textnormal{max}}$ and $\Gamma_{\textnormal{max}}$ and a higher probability to sample the ground state---i.e., $P(0)$ is larger for $p=15$ than for $p=1$. However, while the single layer solution is highly concentrated over the exponential curve associated to a Boltzmann distribution~\cite{DiezValle_PRL23}, the multi-layer QAOA state exhibits a bimodal structure
\begin{equation}\label{eq_probQAOAisBoltzmann}
    P_{\rm QAOA} (E_\zb) \approx B(E_\zb),
\end{equation}
with
\begin{equation}\label{eq_bimodalBoltzmann}
    B(E_\zb) =\frac{1}{\mathcal{Z}_B} \left(c_{\textnormal{high}} e^{-\beta_{\textnormal{high}} E_\zb} + c_{\textnormal{low}} e^{-\beta_{\textnormal{low}} E_\zb}\right),
\end{equation}
where $\beta_{\textnormal{high}}$ and $\beta_{\textnormal{low}}$ ($\beta_{\textnormal{high}}>\beta_{\textnormal{low}}$) correspond to two distinct inverse temperatures characterizing the low- and high-energy (high and low inverse temperature) regime of the spectrum respectively, $c_{\textnormal{high}}$ and $c_{\textnormal{low}}$ are multiplicative factors reflecting the weight of each Boltzmann in the overall probability distribution, and $\mathcal{Z}_B$ is the normalization factor,
\begin{equation}
\mathcal{Z}_B = \sum_\zb \left(c_{\textnormal{high}} e^{-\beta_{\textnormal{high}} E_\zb} + c_{\textnormal{low}} e^{-\beta_{\textnormal{low}} E_\zb}\right).
\label{eq_binomialnormalization}
\end{equation}

To corroborate Eq.~\eqref{eq_probQAOAisBoltzmann}, and calibrate the free parameters $\beta_{\textnormal{high}},\, \beta_{\textnormal{low}},\, c_{\textnormal{high}}$, and $c_{\textnormal{low}}$, we fit the probability distribution $P_\text{QAOA}(E)$ using a modified Kullback–Leibler (KL) divergence
\begin{equation}
   D_{\rm KL}(B||P_\text{QAOA}) = \sum_\zb \mathcal{B}(E_\zb)\log\left[{\frac{B(E_\zb)}{P_\text{QAOA}(E_\zb)}}\right],
   \label{eq_KLdivergence}
\end{equation}
that weights the bimodal Boltzmann distribution $\mathcal{B}(E_\zb) = w(E_\zb) B(E_\zb)$ with a function $w(E_\zb) = 100 E_\zb + 1$. This weight enhances the sensitivity of the fit to the tail of the distribution, particularly at large QAOA depths, where the fraction of states participating in the coldest temperature is very small. As shown in Fig.~\ref{fig_QAOA_distribution}, this weighting captures very well both the high and low energy parts of the spectrum.

\begin{figure}
\includegraphics[width=0.9\linewidth]{ 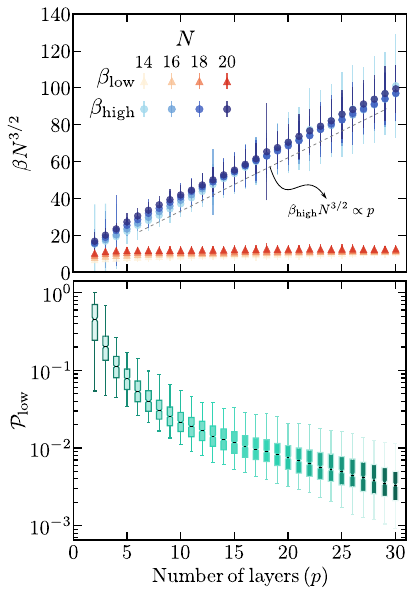}
\caption{\label{fig_QAOAbetafits}(Color online) Evolution of binomial pseudo-Boltzmann states described by Eq.~\eqref{eq_bimodalBoltzmann} with the number of QAOA layers $p$, for 800 random QUBO instances (500 instances for $N=20$). (top) Average effective hot and cold temperatures, together with standard deviation as error bars. (bottom) Distribution of overlaps with the ground state, showing the first to the third quartile (box), the median (solid line) and the 90\% percentile interval (error bars).}
\end{figure}

The minimization of Eq.~\eqref{eq_KLdivergence} for random QSNet instances allows us to analyze the improvement of the QAOA approximation as we increase the number of layers in the circuit. As illustrated in Fig.~\ref{fig_QAOAbetafits}a, the inverse temperature associated to the coldest component $\beta_{\rm high}$ in the wavefunction---the one that concentrates around the ground state---grows linearly with the number of layers $p$, while the hotter noise associated to the tails of the distribution saturates to an asymptotic value. Note that in single layer limit, $p=1$, both temperatures tend to the same value $\beta_{\rm high}=\beta_{\rm low}$, recovering the pseudo-Boltzmann distributions reported for single-layer QAOA~\cite{DiezValle_PRL23,DiezValle_Frontiers2024}.

While the temperature of the noise in the QAOA state remains stationary with increasing number of layers, its contribution to the state decreases exponentially. As shown in Fig.~\ref{fig_QAOAbetafits}b, the relative weight of the tail of the distribution
\begin{equation}
    \mathcal{P}_{\rm low} = 1 - \mathcal{P}_{\rm high},
\end{equation}
given by
\begin{equation}
    \mathcal{P}_{\rm high} = \sum_\zb \frac{1}{\mathcal{Z}_B}\left[{c_{\textnormal{high}} e^{-\beta_{\textnormal{high}} E_\zb}}\right],
\end{equation}
vanishes as $p$ increases. Thus, for large number of layers, the contribution of the coldest component $\beta_{\rm high}$ dominates the global distribution.

This evolution is consistent with the limit of perfect optimization with an infinite number of QAOA layers, in which the probability amplitude becomes zero for any excited state and the system is at temperature zero ($\beta_{\rm high}\rightarrow\infty$). We therefore denote $\beta\equiv\beta_{\rm high}\sim N^{-\frac{3}{2}}$ as the effective temperature of the system that explains the optimization performance of the algorithm at large depth, in agreement with the exponential scaling observed in Ref.~\cite{Lotshaw_PRA23}.

\subsection{Multilayer QAOA gives universal trajectories}
\label{sec_universalAQCtrajectories}
\begin{figure}
\includegraphics[width=0.8\linewidth]{ 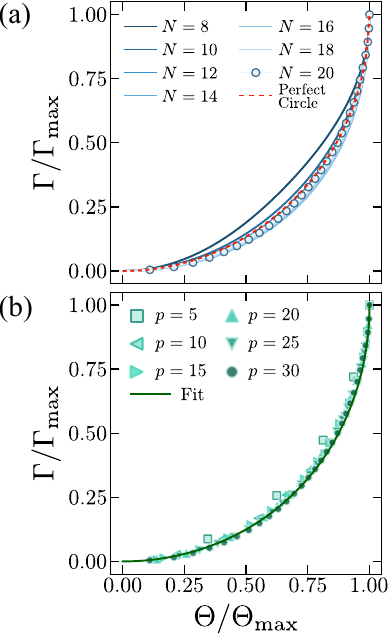}
\caption{(Color online) Annealing trajectories defined by the rescaled optimal QAOA angles $\left(\overline{\Theta}_n/\overline{\Theta}_{\textnormal{max}}, \overline{\Gamma}_n/\overline{\Gamma}_{\textnormal{max}}\right)$. (a) Average trajectories fitted to Eq.~\eqref{eq_AQCcirclefit} for $N=$ 8, 10, 12, 14, 16, 18, and 20 qubits, computed from $p=30$ QAOA angles on 800 QUBO instances (500 for
  $N=$ 20). We also plot the average angles for $N=20$ qubits as circles. (a) Average rescaled optimal angles for QAOA circuits with $N=18$ spins and increasing number of layers $p=$ 5, 10, 15, 20, 25, and 30 layers, computed over 800 QUBO instances, together with a numerical fit to Eq.~\ref{eq_AQCcirclefit} for $p=30$ layers (solid line).}
\label{fig_AQCtrajectoriesfromQAOAangles}
\end{figure}
The previous section analyzed the physical properties of the states generated by optimized multilayer QAOA circuits, running the same optimization protocol over hundreds of random QUBO problems, with up to $N=20$ qubits and up to $p=30$ QAOA layers. The outcome of this study provides us not only with probability distributions of physical interest but also with optimal angles that generate them. Those angles can be analyzed from two different perspectives. We may investigate the integrated angles $\Theta_\text{max}$ and $\Gamma_\text{max}$ as a measure of the resources used by the QAOA protocol to approximate the ground state or to generate pseudo-Boltzmann distributions of specific temperatures. However, we can also analyze the specific paths, investigating the concentration of angles reported in the literature~\cite{Farhi_2022, Basso_quantum_2022, Lee_depth_2023, Zhou_PRX20, Akshay_parameter_2021, Claes_2021}.

Our study focuses on the Hamiltonian paths $(\Theta,\Gamma)$ inferred by the QAOA protocol. The second most important result in this work is the evidence that said paths collapse onto universal curves that approximate annealing paths in the limit of large numbers of layers $p\to\infty$. The collapse of those curves is more evident when one studies separately the rescaled trajectories $(\Theta/\Theta_\text{max}, \Gamma/\Gamma_\text{max})$ from the average total resources  $(\Theta_\text{max},\Gamma_\text{max})$ (c.f. Sec.~\ref{sec_scalingresources}). These rescaled trajectories converge to deformed circles of radius 1, both when we plot different trajectories for increasing number of qubits (c.f. Fig.~\ref{fig_AQCtrajectoriesfromQAOAangles}a), as well as when fix the number of qubits, and we increase the number of layers (c.f. Fig.~\ref{fig_AQCtrajectoriesfromQAOAangles}b).

More precisely, the limit annealing trajectories described by the QAOA angles in Fig.~\ref{fig_AQCtrajectoriesfromQAOAangles} correspond to the polar equation
\begin{equation}\label{eq_AQCcirclefit}
R = 1 + \epsilon(p,N)\left(1 - \cos[4\phi]\right),
\end{equation}
with
\begin{align}
    R^2 (\Theta, \Gamma)&= \left(\frac{\Theta}{\Theta_{\textnormal{max}}}\right)^2 + \left(\frac{\Gamma}{\Gamma_{\textnormal{max}}}\right)^2,\\
    \phi (\Theta, \Gamma) &= \arctan\left[{\frac{ \Theta_{\textnormal{max}}}{ \Gamma_{\textnormal{max}}}\frac{\Gamma}{\Theta}}\right].
\end{align}
In these curves, $\epsilon(p,N)$ is a small parameter characterizing the finite-size deviation of the passage from the radius 1 circle. As shown in Fig.~\ref{fig_AQCtrajectoriesfromQAOAangles}, the QAOA angles distribute evenly along this limit trajectory, with very minor deviations. Indeed, for Fig.~\ref{fig_AQCtrajectoriesfromQAOAangles}a, we find for $N=$ 8, 10, 12, 14, 16, 18, 20, decreasing corrections $\epsilon(N,30)=-0.030,-0.007,0.002,0.014,0.016,0.012,0.005$, with a standard deviation equal to $0.001$. In Appendix~\ref{appen_universaltrajectoryMaxCut}, we show how these results also extend to other models beyond QUBO, such as the maximum cut.

\subsection{Integrated angles and computational cost}
\label{sec_scalingresources}
\begin{figure}
\includegraphics[width=0.75\linewidth]{ 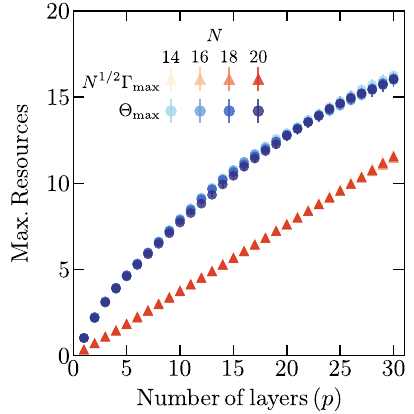}
\caption{(Color online) Scaling of the average total resources $\Gamma_{\textnormal{max}}\sqrt{N}$ (triangles) and $\Theta_{\textnormal{max}}$ (circles) with increasing number of QAOA layers $p$ for different problem sizes. We plot the average of 800 random QUBO instances (500 instances for $N=20$) using a 95\% confidence interval that results indistinguishable from the plot marker. }
\label{fig_AQCtimeVSQAOAlayers}
\end{figure}
In Sec.~\ref{sec:qaoa-with-qa} we introduced the integrated angles $\Theta_\text{max}$ and $\Gamma_\text{max}$ as plausible measures of the computational resources implied by both quantum annealing and the QAOA algorithm. Fig.~\ref{fig_AQCtimeVSQAOAlayers} shows how these numbers increase with the number of layers for different problem sizes. As anticipated before, there is an exceptionally good collapse of the average values over the different problem sizes---illustrated here for $N=14,16,18,20$---with a smooth, monotonous increase of the resources with the number of layers. Remarkably, the integrated angles of the interaction term increase linearly with the number of layers $\Gamma_\text{max}\propto p$ and follows an algebraic decay with the system size, $\Gamma_\text{max}\propto N^{-1/2}$. The mixer contribution, on the other hand follows a trend that also becomes linear in the limit of large number of layers, $\Theta_\text{max}\propto p$, but is independent of the number of qubits, reinforcing the idea that $\Theta_\text{max}$ is a good measure of ``time''. Note that in both cases the growth of the integrated angles with $p$ implies that they are a good proxy for the actual cost of implementing these circuits in a quantum computer, which also increases linearly with the number of layers.

\subsection{Rescaling a trajectory changes the temperature}
\label{sec_rescaling}
Ref.~\cite{DiezValle_PRL23} showed that scaling down the angles of an optimized single layer QAOA circuit---i.e., reducing $\gamma_1$ from the optimal value down to 0---produced pseudo Boltzmann of increasing temperature, making this an effective quantum simulator of partition functions. In our multilayer scenario, the equivalent transformation would be to modify $\Gamma_\text{max}$ or $\Theta_{\max}$ by a constant factor, while preserving the shape of the rescaled trajectory $(\Theta/\Theta_\text{max}, \Gamma/\Gamma_\text{max})$. Traced back to our connection with annealing passages, reducing $\Gamma_{\max}$ by a factor $\zeta$ amounts to speeding up the annealing passage, while keeping the same trajectory
\begin{equation}
    \pmb \gamma_{p} \rightarrow \pmb \gamma_{p}/\zeta = (\gamma_1/\zeta,\ldots,\gamma_p/\zeta).
\end{equation}

\begin{figure}
\includegraphics[width=1.0\linewidth]{ 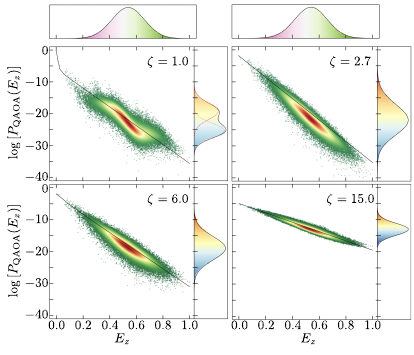}
\caption{(Color online) Effect of rescaling the optimal QAOA trajectories by a factor $\zeta$ $\left(\Gamma_{\textnormal{max}}/\zeta\right)$, for a random QUBO instance with $N=16$ spins and $p=20$ layers. As the trajectory is shortened, the hot component decreases and the distribution approximates better a pseudo-Boltzmann state with a higher temperature.}
\label{fig_QAOAreescaled}
\end{figure}

Fig.~\ref{fig_QAOAreescaled} illustrates the consequences of this rescaling for the quantum state of a single QUBO instance. These plots illustrate the fact that, as we reduce the adiabaticity, the system undergoes a transformation towards an unimodal Boltzmann distribution in which the two modes $\beta_{\rm low}$, $\beta_{\rm high}$ coexist in the range $\zeta=[1,2]$ until the $\mathcal{P}_{\rm high}$ contribution completely cancels out. The coldest state $\beta=\beta_{\rm high}$ is reached for $\zeta=1$ with a contribution of the low beta mode very small $\mathcal{P}_{\rm low}\sim10^{-2}$. On the other side, the least noisy Boltzmann distributions are reached at high temperature for high $\zeta$.
\begin{figure*}
\includegraphics[width=0.85\linewidth]{ 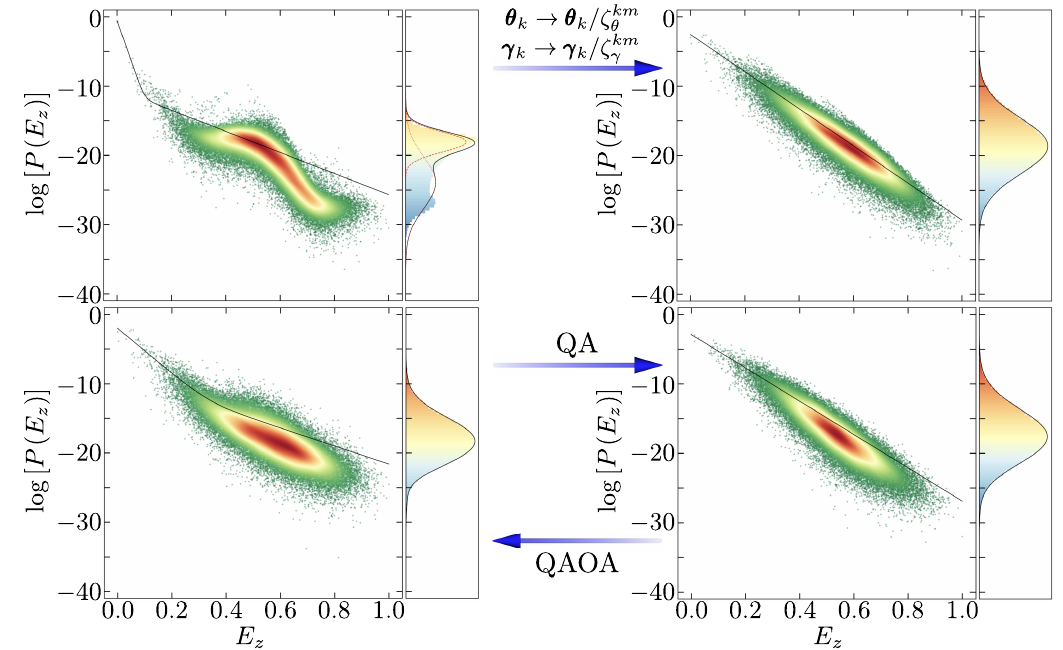}
\caption{\label{fig_QAOAconvergetoQA}(Color online) Convergence of QAOA protocols to quantum annealing passages, and evidence that the effective noise temperature $\beta_{\rm low}$ arises as an effect of the QAOA discretization error. We plot the optimal QAOA output of an $N=16$ instance with $k=25$ (top-left) and $m=5$ (bottom-left) layers. We also show the result rescaling the optimal angles of the $k=25$ instance to the integrated optimal angles of $m=5$ layers (top-right), and the numerical simulation of the continuous QA evolution following the interpolated QAOA trajectory and Eq.~\eqref{eq_aqcshro} (bottom-right).} 
\end{figure*}

\subsection{Multilayer QAOA protocols converge to QA}
The results from Sect.~\ref{sec_universalAQCtrajectories} suggest that QAOA converges towards continuous Hamiltonian trajectories, similar to those that one would expect from QA. This opens three questions: (i) Is the continuous limit of these QAOA trajectories a valid QA path? (ii) Does QA produce the same type of quantum states? (iii) And, if QAOA is approximating a QA path, how can we explain the persistent noise at a hotter temperature?

To answer these questions, let us put forward the following working hypothesis: QAOA acts as a low-order Trotterization of a universal QA path which, when scaled to certain angles $(\Theta_\text{max},\Gamma_\text{max})$, produces pseudo-Boltzmann states at a fixed temperature. In this scenario, the bimodal nature of QAOA states and the $\beta_\text{low}$ is explained mainly by Trotter errors. Consequently, increasing $p$ does not immediately lead to the disappearance of Trotter errors, because when we increase the number of layers the total duration of the evolution $(\Theta_\text{max}^p, \Gamma_\text{max}^p)$ also grows, leading to a finite Trotter error that decreases more slowly with the resources.

To validate this hypothesis we can take a random instance, optimized with $m=5$ layers, as a target temperature. We can then use optimized protocols with $k>m$ layers and rescale their trajectories (c.f. Sec.~\ref{sec_rescaling}) to achieve the same integrated angles and thus the same Hamiltonian path
\begin{equation}
        \pmb \theta_{k} \rightarrow \pmb \theta_{k}/\zeta_{\theta}^{km}\,\,,\,\,\pmb \gamma_{k} \rightarrow \pmb \gamma_{k}/\zeta_{\gamma}^{km}.
\end{equation}
Here $\zeta_{\theta}^{km} = \Theta^k_\text{max}/ \Theta^m_\text{max}$, $\zeta_{\gamma}^{km} = \Gamma^k_{\text{max}}/ \Gamma^m_{\text{max}}$ are the rescaling factors, and $\Theta^k_\text{max},\Gamma^k_\text{max}$ are the integrated angles obtained for QAOA with $k$ layers. We expect that the Trotter errors and deviations from the universal trajectories should be smaller for the $k$-layer than for the $m$-layer experiments. This is verified by simulating the continuous $k\to\infty$ limiting trajectory, rescaled to the same effective resources.

Fig.~\ref{fig_QAOAconvergetoQA} depicts an instance of such a study. We start from two QAOA circuits with $k=25$ layers (top-left) and $m=5$ layers (bottom-left), which produce bimodal distributions with very different temperatures. Our first approach involves rescaling the resources of the $k=25$ circuit to those of the $m=5$ layers protocol. As shown in Fig.~\ref{fig_QAOAconvergetoQA}(top-right), this results in a pseudo-Boltzmann distribution that has a single effective temperature, close to that of the $m=5$ protocol, but without appreciable $\beta_\text{low}$ noise.

We then numerically solved the continuous Schrödinger equation~\eqref{eq_aqcshro} using an approximation to the $k\to\infty$ curve. In particular, we take an interpolation of the converged trajectory of QAOA angles with 30 layers, using it to compute $\Gamma(\Theta)$ and rescaling it to produce an annealing path which reaches up to $(\Theta^5_\text{max},\Gamma^5_\text{max})$. This rescaled universal quantum annealing path is simulated in a numerically exact way. The outcome, shown in Fig.~\ref{fig_QAOAconvergetoQA}(bottom-right), is a single-mode pseudo-Boltzmann distribution. This distribution is very close to the one produced by the rescaled $k=15$ layers experiment in Fig.~\ref{fig_QAOAconvergetoQA}(top-right).

We take this result as a confirmation of our hypothesis---i.e., the bimodal nature of QAOA protocols can be explained by the Trotterization errors of QA trajectories. Furthermore, these simulations also suggest that QA and QAOA have similar behaviors as cooling protocols with limited temperatures determined by the Hamiltonian paths. In this picture, imperfections in QA protocols are also explained as pseudo-Boltzmann distributions or ``heating'' that decreases with the QA passage length. This behavior is somewhat contradictory to earlier QA pictures based on the closing of gaps and phase transitions along the annealing passage. However, the emergence of these states could be explained at a higher level, assuming that QA and QAOA errors experience some kind of detailed-balance relation, whereby the rate of excitation and cooling are dictated by the protocol duration or ``annealing speed''.

\subsection{Scaling of QAOA resources}
The problem with the plot in Fig.~\ref{fig_AQCtimeVSQAOAlayers} is that it only reflects the cost of making larger circuits, but not how much the solution ``improves'' with those resources. However, in order to make clear statements about the performance of the algorithm, we need to understand the growth of the cost with the solution quality and the problem size. From a computational perspective, scaling refers to the relationship between the resource requirements of an algorithm and the growth of the problem size, as well as its impact on performance.

In quantum annealing, where the focus lies on preparing good ground states, a more canonical plot would be to study the \textit{time-to-solution}, that is, what resources are needed to achieve a fixed overlap with the target state. Fortunately, in our study we already have a very general figure of merit---$\beta_\text{high}$, the temperature of the produced states (c.f. Fig.~\ref{fig_QAOAbetafits})---which is not a prefixed tolerance, but a property of the algorithm's outcome.

\begin{figure}[t]
\includegraphics[width=0.8\linewidth]{ 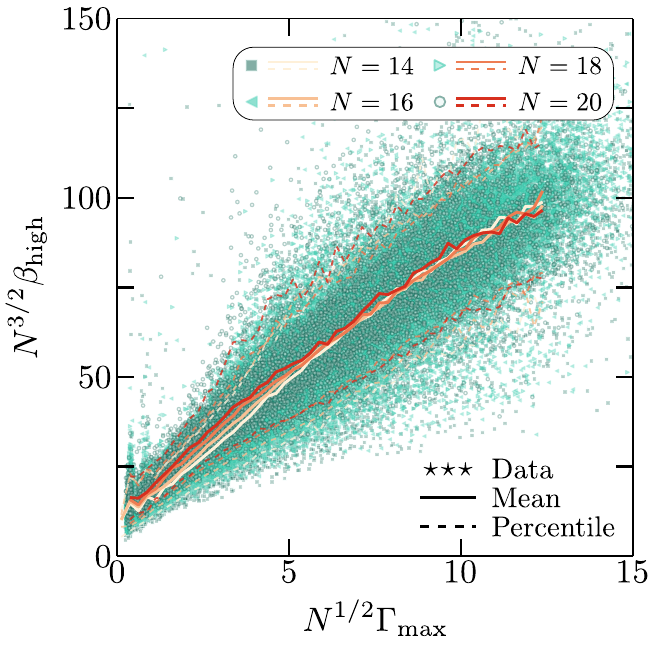}
\caption{(Color online) Evolution of the effective temperature $\beta\equiv\beta_{\rm high}$ with QA resources $\Gamma_{\textnormal{max}}$ extrapolated from optimal QAOA circuits of $p=[1,...,30]$. We plot the results of 800 (500 for N=20) random QUBO instances of $N=$ 14, 16, 18, 20 spins (markers), together with the average of these results (solid lines), and the 90\% percentile intervals (dashed lines). }
\label{fig_QAOAbetavsresources}
\end{figure}

To explore the connection between the quality of the solutions and the integrated resources, Fig.~\ref{fig_QAOAbetavsresources} plots the results of simulating hundreds of QUBO problems for instances with $N=14, 16, 18$ and 20 qubits. The plot shows one point for each random problem, plotting the effective inverse temperature $\beta_\text{high}$ of the resulting state vs. the integrated angle $\Gamma_\text{max}$. The plot shows that inverse temperature of the QAOA states scales polynomially with the size of the system $\beta_\text{high}\propto N^{-\frac{3}{2}}$, and linearly with the quantum circuits depth $\beta_\text{high}\propto \Gamma_\text{max}\propto p$. These are promising results, suggesting that the QAOA protocol is a very efficient cooling method with resources that scale in a way that is advantageous for most random problems---Fig.~\ref{fig_QAOAbetavsresources} includes 90\% percentile intervals.

\section{Conclusions and outlook}
\label{sec_conclusions}
In this work we have performed a physical study of the properties of optimized QAOA circuits and their connection to quantum annealing trajectories. We have established that there exists a formal connection between QAOA and QA, as both explore paths in Hamiltonian space. This connection becomes empirically exact, as the average angles of QAOA circuits collapse over universal trajectories - defined on the family of dense QUBO problems - that approach quantum annealing paths in the limit of infinite many layers. These optimal trajectories have been shown to produce, both in the QAOA and QA scenarios, similar pseudo-Boltzmann distributions---these are states whose probability distribution approximates an exponential $\exp(-\beta_\text{high} E)$ but have scrambled phases---. However, QAOA protocols exhibit an additional background noise that can be explained as the Trotter errors that result from approximating QA trajectories. In both cases, the coldest temperature of these distributions exhibits favorable scalings with respect to the problem size and the integrated resources, which in QAOA also correspond to the number of layers in the algorithm, $T\sim\mathcal{O}(p^{-1})$. However, both QAOA and QA paths can be used to produce hotter states simply by rescaling the Hamiltonian trajectories to shorter paths than the optimal one.

This work provides a unifying picture of QAOA and quantum annealing, connecting earlier results on the concentration of parameters~\cite{Farhi_2022, Basso_quantum_2022, Lee_depth_2023, Zhou_PRX20, Akshay_parameter_2021, Claes_2021} as well as with recent studies that derive QAOA trajectories from QA paths~\cite{boulebnane2025}. However, multiple questions remain still open along this connection. In the QAOA domain we have not addressed how to use the discovered universal trajectories in actual cooling protocols for unknown problems, and we don't have yet an explanation for the nature of these trajectories. More interestingly, this work has revealed that QAOA is a viable tool to design annealing paths, that avoids the need to study precisely the eigenstate spectrum of the deformed Hamiltonian $\hat{H}(s)$. It remains to study the quality of these paths in large-scale annealing protocols and bigger problems, a study that is much more costly than the finite-size circuit simulations we performed in this work.

Our study has also confirmed the utility of QAOA circuits and QA passages as simulators of partition functions and thermal distributions. Not only this thermal characterization of the global output distribution has shown a valid metric for the performance of QAOA algorithm~\cite{Lotshaw_PRA23}, but we have shown that QAOA and QA protocols produce similar distributions. This seems to contradict the folklore picture that assumes QA to have a catastrophic performance for moderate speeds, suggesting instead that annealing errors behave like thermal noise, whose temperature can be controlled with the annealing speed. It would be both very interesting to verify this hypothesis with larger simulations and study the resilience of this picture in real hardware with other sources of noise. Indeed, while our performance metrics suggest a favorable behavior of QAOA on most instances of random problems that include spin glasses, it remains to confirm the nature of this scaling and its robustness in practical scenarios, including a broader set of problems with structured and sparse connectivity.

Finally, the utility of QAOA for thermal state simulation has applications beyond the approximation of NP-hard problems. The sampling of pseudo-Boltzmann distributions could be a useful subroutine for physical simulation, combinatorial optimization, or machine learning. In this sense, it would be interesting to verify the tunability of said simulators in real hardware, the comparison to alternative methods, as well as their extension to other types of ansätze or Hamiltonians in line with recent results~\cite{chen_effective_2024}.

\begin{acknowledgments}
This work has been supported by European Commission FET Open project AVaQus Grant Agreement No. 899561, Comunidad de Madrid Sinergicos 2020 project NanoQuCo-CM (Y2020/TCS- 6545), CSIC Quantum Technologies Platform PTI-001, Spanish project PID2021-127968NBI00 funded by MICIU/AEI/10.13039/501100011033, 
and the Ministry for Digital Transformation and of Civil Service of the Spanish Government through the QUANTUM
ENIA project call - Quantum Spain project, and by the European Union through
the Recovery, Transformation and Resilience Plan - NextGenerationEU within
the framework of the Digital Spain 2026 Agenda. FJGR acknowledges financial support from the Spanish Government via the project PID2024-161371NB-C21 (MCIU/AEI/FEDER, EU).
\end{acknowledgments}

\appendix

\section{Numerical details}

In this appendix we provide additional details about the numerical experiments that support the results reported in the manuscript. 

All quantum circuits were simulated using exact state-vector calculations under zero-noise conditions. The optimization of the variational parameters involved in the numerical results of QAOA at any depth was performed using the L-BFGS-B method implemented in scipy~\cite{2020SciPy-NMeth}. The gradients were calculated analytically, and the convergence criteria established that the optimization stops when 
\begin{equation}
\frac{f^k - f^{k+1}}{\max(|f^k|,|f^{k+1}|,1)} \leq f_{tol} = 2\cdot10^{-9},
\end{equation}
or 
\begin{equation}
\max{|\textnormal{proj(} g_i) |\; i = 1, ..., n} <= 10^{-10},    
\end{equation}
with $f^k$ the value of the cost function at iteration $k$ and $\textnormal{proj(} g_i)$ the i-th component of the projected gradient. The parameters of a QAOA circuit with $p$ layers are initialized using the optimal parameters of $p-1$,   
\begin{equation}
    \theta^p_l = \Theta^{p-1}_{\max}/p \;\;\;,\;\;\;\gamma^p_l = \Gamma^{p-1}_{\max}/p,
\end{equation}
for $l = \{1,...,p\}$, with $\theta^1 = \pi/3$ and $\gamma^1 = 0.1$. This heuristic was tested in different scenarios and provided good results in practice.

As for the simulation of continuous quantum annealing passages (see Fig.\ref{fig_QAOAconvergetoQA}), we numerically integrated the system of ordinary differential equations given by the normalized continuous Schrödinger equation between $\Theta=0$ and $\Theta=1$,
\begin{equation}
  \partial_\Theta \ket{\psi(\Theta)} = -i\left[\Gamma_{\max} \frac{d\Gamma}{d\Theta} \hat{H}_{{\rm QSNet}} - \Theta_{\max} \hat{H}_x \right]\ket{\psi(\Theta)}.
  \label{eq_aqcshro}
\end{equation}
The integration method used was the explicit Runge-Kutta method of order 5(4) implemented in scipy~\cite{2020SciPy-NMeth} with a maximum step size equal to $10^{-5}$. \\

\section{Universal trajectories on MaxCut}
\label{appen_universaltrajectoryMaxCut}

\begin{figure*}
\includegraphics[width=0.8\linewidth]{ 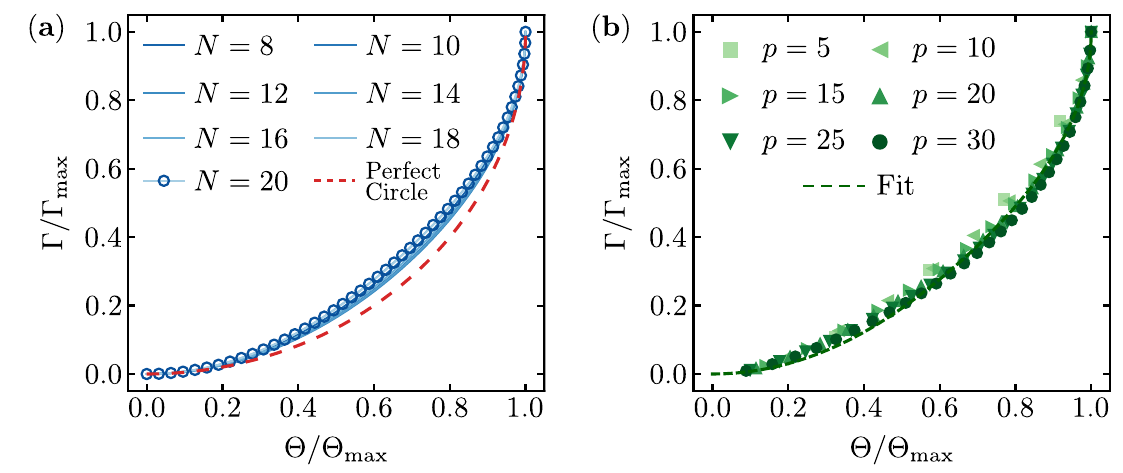}
\caption{(Color online) Annealing trajectories defined by the rescaled optimal QAOA angles $\left(\overline{\Theta}_n/\overline{\Theta}_{\textnormal{max}}, \overline{\Gamma}_n/\overline{\Gamma}_{\textnormal{max}}\right)$. (a) Average trajectories fitted to Eq.~\eqref{eq_AQCcirclefit} for $N=$ 8, 10, 12, 14, 16, 18, and 20 qubits, computed from $p=30$ QAOA angles on 800 MaxCut instances (500 for
  $N=$ 20). We also plot the average angles for $N=20$ qubits as circles. (a) Average rescaled optimal angles for QAOA circuits with $N=18$ spins and increasing number of layers $p=$ 5, 10, 15, 20, 25, and 30 layers, computed over 800 MaxCut instances, together with a numerical fit to Eq.~\ref{eq_AQCcirclefit} for $p=30$ layers (solid line).}
\label{fig_AQCtrajectoriesfromQAOAangles_MAXCUT}
\end{figure*}

The figures presented in the main text show universal behaviors of randomly generated QUBO problems~\eqref{EQubo}. In this appendix, we extend the scope of the results to a different family of NP-hard binary combinatorial optimization problems, the maximum cut problem (MaxCut). MaxCut is defined by the following classical energy:
\begin{equation}
E_{{\rm MaxCut}}(\xb) =-2\sum_{n,m=1}^{N}x_{n}Q_{n,m}(1-x_m).
\label{EMaxCut}
\end{equation}
where $\mathbf{\xb} = (x_1,\ldots,x_N)$ with $x_i \in \{0,1\}$, and $\pmb Q$ is an $N$-by-$N$ square symmetric matrix of real coefficients denoted by $Q_{nm}$. As in QUBO instances, we consider fully connected graphs, i.e. all off-diagonal elements of $\mathbf{Q}$ or $\mathbf{J}$ are nonzero values randomly drawn from a normal distribution $\mathcal{N}(\mu,\sigma^2)$ with mean $\mu = 0$ and variance $\sigma^2 =1$. Although MaxCut~\eqref{EMaxCut} can be formulated as a QUBO~\eqref{EQubo} by allowing diagonal elements in the Q matrix, MaxCut exhibits a global symmetry $\mathbb{Z}_2$ generally absent in QUBO instances.

Fig.\ref{fig_AQCtrajectoriesfromQAOAangles_MAXCUT} shows that the optimal QAOA angles on MaxCut instances collapse into universal curves. As was the case for the QUBO problem class, these curves define annealing paths described by deformed radius 1 circles given by equation~\eqref{eq_AQCcirclefit}. Specifically, we find small corrections $\epsilon(N,p=30)=-0.026\pm0.001,-0.019\pm0.001,-0.017\pm0.001,-0.019\pm0.001,-0.022\pm0.002,-0.031\pm0.002,-0.031\pm0.002$ for $N=$ 8, 10, 12, 14, 16, 18, 20, respectively.  

\color{black}



\bibliography{My_bib_QAOA}

\begin{thebibliography}{38}%
\makeatletter
\providecommand \@ifxundefined [1]{%
 \@ifx{#1\undefined}
}%
\providecommand \@ifnum [1]{%
 \ifnum #1\expandafter \@firstoftwo
 \else \expandafter \@secondoftwo
 \fi
}%
\providecommand \@ifx [1]{%
 \ifx #1\expandafter \@firstoftwo
 \else \expandafter \@secondoftwo
 \fi
}%
\providecommand \natexlab [1]{#1}%
\providecommand \enquote  [1]{``#1''}%
\providecommand \bibnamefont  [1]{#1}%
\providecommand \bibfnamefont [1]{#1}%
\providecommand \citenamefont [1]{#1}%
\providecommand \href@noop [0]{\@secondoftwo}%
\providecommand \href [0]{\begingroup \@sanitize@url \@href}%
\providecommand \@href[1]{\@@startlink{#1}\@@href}%
\providecommand \@@href[1]{\endgroup#1\@@endlink}%
\providecommand \@sanitize@url [0]{\catcode `\\12\catcode `\$12\catcode
  `\&12\catcode `\#12\catcode `\^12\catcode `\_12\catcode `\%12\relax}%
\providecommand \@@startlink[1]{}%
\providecommand \@@endlink[0]{}%
\providecommand \url  [0]{\begingroup\@sanitize@url \@url }%
\providecommand \@url [1]{\endgroup\@href {#1}{\urlprefix }}%
\providecommand \urlprefix  [0]{URL }%
\providecommand \Eprint [0]{\href }%
\providecommand \doibase [0]{https://doi.org/}%
\providecommand \selectlanguage [0]{\@gobble}%
\providecommand \bibinfo  [0]{\@secondoftwo}%
\providecommand \bibfield  [0]{\@secondoftwo}%
\providecommand \translation [1]{[#1]}%
\providecommand \BibitemOpen [0]{}%
\providecommand \bibitemStop [0]{}%
\providecommand \bibitemNoStop [0]{.\EOS\space}%
\providecommand \EOS [0]{\spacefactor3000\relax}%
\providecommand \BibitemShut  [1]{\csname bibitem#1\endcsname}%
\let\auto@bib@innerbib\@empty
\bibitem [{\citenamefont {Aharonov}\ \emph {et~al.}(2004)\citenamefont
  {Aharonov}, \citenamefont {{van Dam}}, \citenamefont {Kempe}, \citenamefont
  {Landau}, \citenamefont {Lloyd},\ and\ \citenamefont {Regev}}]{Aharonov04}%
  \BibitemOpen
  \bibfield  {author} {\bibinfo {author} {\bibfnamefont {D.}~\bibnamefont
  {Aharonov}}, \bibinfo {author} {\bibfnamefont {W.}~\bibnamefont {{van Dam}}},
  \bibinfo {author} {\bibfnamefont {J.}~\bibnamefont {Kempe}}, \bibinfo
  {author} {\bibfnamefont {Z.}~\bibnamefont {Landau}}, \bibinfo {author}
  {\bibfnamefont {S.}~\bibnamefont {Lloyd}},\ and\ \bibinfo {author}
  {\bibfnamefont {O.}~\bibnamefont {Regev}},\ }\bibfield  {title} {\bibinfo
  {title} {Adiabatic quantum computation is equivalent to standard quantum
  computation},\ }in\ \href {https://doi.org/10.1109/FOCS.2004.8} {\emph
  {\bibinfo {booktitle} {45th Annual {{IEEE}} Symposium on Foundations of
  Computer Science}}}\ (\bibinfo {year} {2004})\ pp.\ \bibinfo {pages}
  {42--51}\BibitemShut {NoStop}%
\bibitem [{\citenamefont {Albash}\ and\ \citenamefont
  {Lidar}(2018)}]{Lidar_RMP18}%
  \BibitemOpen
  \bibfield  {author} {\bibinfo {author} {\bibfnamefont {T.}~\bibnamefont
  {Albash}}\ and\ \bibinfo {author} {\bibfnamefont {D.~A.}\ \bibnamefont
  {Lidar}},\ }\bibfield  {title} {\bibinfo {title} {Adiabatic quantum
  computation},\ }\href {https://doi.org/10.1103/RevModPhys.90.015002}
  {\bibfield  {journal} {\bibinfo  {journal} {Reviews of Modern Physics}\
  }\textbf {\bibinfo {volume} {90}},\ \bibinfo {pages} {015002} (\bibinfo
  {year} {2018})}\BibitemShut {NoStop}%
\bibitem [{\citenamefont {Farhi}\ \emph
  {et~al.}(2014{\natexlab{a}})\citenamefont {Farhi}, \citenamefont
  {Goldstone},\ and\ \citenamefont {Gutmann}}]{Farhi14A}%
  \BibitemOpen
  \bibfield  {author} {\bibinfo {author} {\bibfnamefont {E.}~\bibnamefont
  {Farhi}}, \bibinfo {author} {\bibfnamefont {J.}~\bibnamefont {Goldstone}},\
  and\ \bibinfo {author} {\bibfnamefont {S.}~\bibnamefont {Gutmann}},\ }\href
  {https://doi.org/10.48550/ARXIV.1411.4028} {\bibinfo {title} {A quantum
  approximate optimization algorithm}} (\bibinfo {year} {2014}{\natexlab{a}}),\
  \Eprint {https://arxiv.org/abs/1411.4028} {arXiv:1411.4028 [quant-ph]}
  \BibitemShut {NoStop}%
\bibitem [{\citenamefont {Farhi}\ \emph
  {et~al.}(2014{\natexlab{b}})\citenamefont {Farhi}, \citenamefont
  {Goldstone},\ and\ \citenamefont {Gutmann}}]{Farhi14B}%
  \BibitemOpen
  \bibfield  {author} {\bibinfo {author} {\bibfnamefont {E.}~\bibnamefont
  {Farhi}}, \bibinfo {author} {\bibfnamefont {J.}~\bibnamefont {Goldstone}},\
  and\ \bibinfo {author} {\bibfnamefont {S.}~\bibnamefont {Gutmann}},\ }\href
  {https://doi.org/10.48550/ARXIV.1412.6062} {\bibinfo {title} {A quantum
  approximate optimization algorithm applied to a bounded occurrence constraint
  problem}} (\bibinfo {year} {2014}{\natexlab{b}}),\ \Eprint
  {https://arxiv.org/abs/1412.6062} {arXiv:1412.6062 [quant-ph]} \BibitemShut
  {NoStop}%
\bibitem [{\citenamefont {Farhi}\ \emph
  {et~al.}(2022{\natexlab{a}})\citenamefont {Farhi}, \citenamefont {Goldstone},
  \citenamefont {Gutmann},\ and\ \citenamefont {Zhou}}]{Farhi_Q22}%
  \BibitemOpen
  \bibfield  {author} {\bibinfo {author} {\bibfnamefont {E.}~\bibnamefont
  {Farhi}}, \bibinfo {author} {\bibfnamefont {J.}~\bibnamefont {Goldstone}},
  \bibinfo {author} {\bibfnamefont {S.}~\bibnamefont {Gutmann}},\ and\ \bibinfo
  {author} {\bibfnamefont {L.}~\bibnamefont {Zhou}},\ }\bibfield  {title}
  {\bibinfo {title} {The {{Quantum Approximate Optimization Algorithm}} and the
  {{Sherrington-Kirkpatrick Model}} at {{Infinite Size}}},\ }\href
  {https://doi.org/10.22331/q-2022-07-07-759} {\bibfield  {journal} {\bibinfo
  {journal} {Quantum}\ }\textbf {\bibinfo {volume} {6}},\ \bibinfo {pages}
  {759} (\bibinfo {year} {2022}{\natexlab{a}})}\BibitemShut {NoStop}%
\bibitem [{\citenamefont {Susa}\ and\ \citenamefont
  {Nishimori}(2021)}]{Susa_PRA21}%
  \BibitemOpen
  \bibfield  {author} {\bibinfo {author} {\bibfnamefont {Y.}~\bibnamefont
  {Susa}}\ and\ \bibinfo {author} {\bibfnamefont {H.}~\bibnamefont
  {Nishimori}},\ }\bibfield  {title} {\bibinfo {title} {Variational
  optimization of the quantum annealing schedule for the
  {L}echner-{H}auke-{Z}oller scheme},\ }\href
  {https://doi.org/10.1103/PhysRevA.103.022619} {\bibfield  {journal} {\bibinfo
   {journal} {Phys. Rev. A}\ }\textbf {\bibinfo {volume} {103}},\ \bibinfo
  {pages} {022619} (\bibinfo {year} {2021})}\BibitemShut {NoStop}%
\bibitem [{\citenamefont {Fin\ifmmode~\check{z}\else \v{z}\fi{}gar}\ \emph
  {et~al.}(2024)\citenamefont {Fin\ifmmode~\check{z}\else \v{z}\fi{}gar},
  \citenamefont {Schuetz}, \citenamefont {Brubaker}, \citenamefont
  {Nishimori},\ and\ \citenamefont {Katzgraber}}]{Helmut_PRR24}%
  \BibitemOpen
  \bibfield  {author} {\bibinfo {author} {\bibfnamefont {J.~R.}\ \bibnamefont
  {Fin\ifmmode~\check{z}\else \v{z}\fi{}gar}}, \bibinfo {author} {\bibfnamefont
  {M.~J.~A.}\ \bibnamefont {Schuetz}}, \bibinfo {author} {\bibfnamefont
  {J.~K.}\ \bibnamefont {Brubaker}}, \bibinfo {author} {\bibfnamefont
  {H.}~\bibnamefont {Nishimori}},\ and\ \bibinfo {author} {\bibfnamefont
  {H.~G.}\ \bibnamefont {Katzgraber}},\ }\bibfield  {title} {\bibinfo {title}
  {Designing quantum annealing schedules using bayesian optimization},\ }\href
  {https://doi.org/10.1103/PhysRevResearch.6.023063} {\bibfield  {journal}
  {\bibinfo  {journal} {Phys. Rev. Res.}\ }\textbf {\bibinfo {volume} {6}},\
  \bibinfo {pages} {023063} (\bibinfo {year} {2024})}\BibitemShut {NoStop}%
\bibitem [{\citenamefont {Chen}\ \emph {et~al.}(2022)\citenamefont {Chen},
  \citenamefont {Chen}, \citenamefont {Lee}, \citenamefont {Zhang},\ and\
  \citenamefont {Hsieh}}]{Chen_NatMach2022}%
  \BibitemOpen
  \bibfield  {author} {\bibinfo {author} {\bibfnamefont {Y.-Q.}\ \bibnamefont
  {Chen}}, \bibinfo {author} {\bibfnamefont {Y.}~\bibnamefont {Chen}}, \bibinfo
  {author} {\bibfnamefont {C.-K.}\ \bibnamefont {Lee}}, \bibinfo {author}
  {\bibfnamefont {S.}~\bibnamefont {Zhang}},\ and\ \bibinfo {author}
  {\bibfnamefont {C.-Y.}\ \bibnamefont {Hsieh}},\ }\bibfield  {title} {\bibinfo
  {title} {Optimizing quantum annealing schedules with {M}onte {C}arlo tree
  search enhanced with neural networks},\ }\href
  {https://doi.org/10.1038/s42256-022-00446-y} {\bibfield  {journal} {\bibinfo
  {journal} {Nat. Mach. Intell.}\ }\textbf {\bibinfo {volume} {4}},\ \bibinfo
  {pages} {269} (\bibinfo {year} {2022})}\BibitemShut {NoStop}%
\bibitem [{\citenamefont {Pelofske}\ \emph {et~al.}(2023)\citenamefont
  {Pelofske}, \citenamefont {B{\"a}rtschi},\ and\ \citenamefont
  {Eidenbenz}}]{Pelofske_2023}%
  \BibitemOpen
  \bibfield  {author} {\bibinfo {author} {\bibfnamefont {E.}~\bibnamefont
  {Pelofske}}, \bibinfo {author} {\bibfnamefont {A.}~\bibnamefont
  {B{\"a}rtschi}},\ and\ \bibinfo {author} {\bibfnamefont {S.}~\bibnamefont
  {Eidenbenz}},\ }\bibfield  {title} {\bibinfo {title} {Quantum annealing vs.
  {{QAOA}}: 127 {Q}ubit {H}igher-{O}rder {I}sing problems on {{NISQ}}
  computers},\ }in\ \href {https://doi.org/10.1007/978-3-031-32041-5_13} {\emph
  {\bibinfo {booktitle} {High Performance Computing}}}\ (\bibinfo  {publisher}
  {Springer Nature Switzerland},\ \bibinfo {year} {2023})\ pp.\ \bibinfo
  {pages} {240--258}\BibitemShut {NoStop}%
\bibitem [{\citenamefont {Pelofske}\ \emph {et~al.}(2024)\citenamefont
  {Pelofske}, \citenamefont {B{\"a}rtschi}, \citenamefont {Cincio},
  \citenamefont {Golden},\ and\ \citenamefont {Eidenbenz}}]{Pelofske2024}%
  \BibitemOpen
  \bibfield  {author} {\bibinfo {author} {\bibfnamefont {E.}~\bibnamefont
  {Pelofske}}, \bibinfo {author} {\bibfnamefont {A.}~\bibnamefont
  {B{\"a}rtschi}}, \bibinfo {author} {\bibfnamefont {L.}~\bibnamefont
  {Cincio}}, \bibinfo {author} {\bibfnamefont {J.}~\bibnamefont {Golden}},\
  and\ \bibinfo {author} {\bibfnamefont {S.}~\bibnamefont {Eidenbenz}},\
  }\bibfield  {title} {\bibinfo {title} {Scaling whole-chip {{QAOA}} for
  higher-order {I}sing spin glass models on heavy-hex graphs},\ }\bibfield
  {journal} {\bibinfo  {journal} {npj Quantum Information}\ }\textbf {\bibinfo
  {volume} {10}},\ \href {https://doi.org/10.1038/s41534-024-00906-w}
  {10.1038/s41534-024-00906-w} (\bibinfo {year} {2024})\BibitemShut {NoStop}%
\bibitem [{\citenamefont {He}\ \emph {et~al.}(2024)\citenamefont {He},
  \citenamefont {Amaro}, \citenamefont {Shaydulin},\ and\ \citenamefont
  {Pistoia}}]{he_2024}%
  \BibitemOpen
  \bibfield  {author} {\bibinfo {author} {\bibfnamefont {Z.}~\bibnamefont
  {He}}, \bibinfo {author} {\bibfnamefont {D.}~\bibnamefont {Amaro}}, \bibinfo
  {author} {\bibfnamefont {R.}~\bibnamefont {Shaydulin}},\ and\ \bibinfo
  {author} {\bibfnamefont {M.}~\bibnamefont {Pistoia}},\ }\href@noop {}
  {\bibinfo {title} {Performance of quantum approximate optimization with
  quantum error detection}} (\bibinfo {year} {2024}),\ \Eprint
  {https://arxiv.org/abs/2409.12104} {arXiv:2409.12104 [quant-ph]} \BibitemShut
  {NoStop}%
\bibitem [{\citenamefont {Pagano}\ \emph {et~al.}(2020)\citenamefont {Pagano},
  \citenamefont {Bapat}, \citenamefont {Becker}, \citenamefont {Collins},
  \citenamefont {De}, \citenamefont {Hess}, \citenamefont {Kaplan},
  \citenamefont {Kyprianidis}, \citenamefont {Tan}, \citenamefont {Baldwin},
  \citenamefont {Brady}, \citenamefont {Deshpande}, \citenamefont {Liu},
  \citenamefont {Jordan}, \citenamefont {Gorshkov},\ and\ \citenamefont
  {Monroe}}]{Monroe_PNAS20}%
  \BibitemOpen
  \bibfield  {author} {\bibinfo {author} {\bibfnamefont {G.}~\bibnamefont
  {Pagano}}, \bibinfo {author} {\bibfnamefont {A.}~\bibnamefont {Bapat}},
  \bibinfo {author} {\bibfnamefont {P.}~\bibnamefont {Becker}}, \bibinfo
  {author} {\bibfnamefont {K.~S.}\ \bibnamefont {Collins}}, \bibinfo {author}
  {\bibfnamefont {A.}~\bibnamefont {De}}, \bibinfo {author} {\bibfnamefont
  {P.~W.}\ \bibnamefont {Hess}}, \bibinfo {author} {\bibfnamefont {H.~B.}\
  \bibnamefont {Kaplan}}, \bibinfo {author} {\bibfnamefont {A.}~\bibnamefont
  {Kyprianidis}}, \bibinfo {author} {\bibfnamefont {W.~L.}\ \bibnamefont
  {Tan}}, \bibinfo {author} {\bibfnamefont {C.}~\bibnamefont {Baldwin}},
  \bibinfo {author} {\bibfnamefont {L.~T.}\ \bibnamefont {Brady}}, \bibinfo
  {author} {\bibfnamefont {A.}~\bibnamefont {Deshpande}}, \bibinfo {author}
  {\bibfnamefont {F.}~\bibnamefont {Liu}}, \bibinfo {author} {\bibfnamefont
  {S.}~\bibnamefont {Jordan}}, \bibinfo {author} {\bibfnamefont {A.~V.}\
  \bibnamefont {Gorshkov}},\ and\ \bibinfo {author} {\bibfnamefont
  {C.}~\bibnamefont {Monroe}},\ }\bibfield  {title} {\bibinfo {title} {Quantum
  approximate optimization of the long-range {{Ising}} model with a trapped-ion
  quantum simulator},\ }\href {https://doi.org/10.1073/pnas.2006373117}
  {\bibfield  {journal} {\bibinfo  {journal} {Proceedings of the National
  Academy of Sciences}\ }\textbf {\bibinfo {volume} {117}},\ \bibinfo {pages}
  {25396} (\bibinfo {year} {2020})}\BibitemShut {NoStop}%
\bibitem [{\citenamefont {Egger}\ \emph {et~al.}(2021)\citenamefont {Egger},
  \citenamefont {Mare{\v c}ek},\ and\ \citenamefont
  {Woerner}}]{Egger_warm_2021}%
  \BibitemOpen
  \bibfield  {author} {\bibinfo {author} {\bibfnamefont {D.~J.}\ \bibnamefont
  {Egger}}, \bibinfo {author} {\bibfnamefont {J.}~\bibnamefont {Mare{\v
  c}ek}},\ and\ \bibinfo {author} {\bibfnamefont {S.}~\bibnamefont {Woerner}},\
  }\bibfield  {title} {\bibinfo {title} {Warm-starting quantum optimization},\
  }\href {https://doi.org/10.22331/q-2021-06-17-479} {\bibfield  {journal}
  {\bibinfo  {journal} {Quantum}\ }\textbf {\bibinfo {volume} {5}},\ \bibinfo
  {pages} {479} (\bibinfo {year} {2021})}\BibitemShut {NoStop}%
\bibitem [{\citenamefont {Tate}\ and\ \citenamefont
  {Eidenbenz}(2024)}]{tate2024theoreticalapproximationratioswarmstarted}%
  \BibitemOpen
  \bibfield  {author} {\bibinfo {author} {\bibfnamefont {R.}~\bibnamefont
  {Tate}}\ and\ \bibinfo {author} {\bibfnamefont {S.}~\bibnamefont
  {Eidenbenz}},\ }\href@noop {} {\bibinfo {title} {Theoretical approximation
  ratios for warm-started {{QAOA}} on 3-regular {M}ax-{C}ut instances at depth
  $p=1$}} (\bibinfo {year} {2024}),\ \Eprint {https://arxiv.org/abs/2402.12631}
  {arXiv:2402.12631 [quant-ph]} \BibitemShut {NoStop}%
\bibitem [{\citenamefont {Lyngfelt}\ and\ \citenamefont
  {Garc\'{\i}a-\'Alvarez}(2025)}]{lyngfelt2025symmetry}%
  \BibitemOpen
  \bibfield  {author} {\bibinfo {author} {\bibfnamefont {I.}~\bibnamefont
  {Lyngfelt}}\ and\ \bibinfo {author} {\bibfnamefont {L.}~\bibnamefont
  {Garc\'{\i}a-\'Alvarez}},\ }\bibfield  {title} {\bibinfo {title}
  {Symmetry-informed transferability of optimal parameters in the quantum
  approximate optimization algorithm},\ }\href
  {https://doi.org/10.1103/PhysRevA.111.022418} {\bibfield  {journal} {\bibinfo
   {journal} {Phys. Rev. A}\ }\textbf {\bibinfo {volume} {111}},\ \bibinfo
  {pages} {022418} (\bibinfo {year} {2025})}\BibitemShut {NoStop}%
\bibitem [{\citenamefont {Zhou}\ \emph {et~al.}(2020)\citenamefont {Zhou},
  \citenamefont {Wang}, \citenamefont {Choi}, \citenamefont {Pichler},\ and\
  \citenamefont {Lukin}}]{Zhou_PRX20}%
  \BibitemOpen
  \bibfield  {author} {\bibinfo {author} {\bibfnamefont {L.}~\bibnamefont
  {Zhou}}, \bibinfo {author} {\bibfnamefont {S.-T.}\ \bibnamefont {Wang}},
  \bibinfo {author} {\bibfnamefont {S.}~\bibnamefont {Choi}}, \bibinfo {author}
  {\bibfnamefont {H.}~\bibnamefont {Pichler}},\ and\ \bibinfo {author}
  {\bibfnamefont {M.~D.}\ \bibnamefont {Lukin}},\ }\bibfield  {title} {\bibinfo
  {title} {Quantum approximate optimization algorithm: {{Performance}},
  mechanism, and implementation on near-term devices},\ }\href
  {https://doi.org/10.1103/PhysRevX.10.021067} {\bibfield  {journal} {\bibinfo
  {journal} {Physical Review X}\ }\textbf {\bibinfo {volume} {10}},\ \bibinfo
  {pages} {021067} (\bibinfo {year} {2020})}\BibitemShut {NoStop}%
\bibitem [{\citenamefont {Lee}\ \emph {et~al.}(2023)\citenamefont {Lee},
  \citenamefont {Xie}, \citenamefont {Cai}, \citenamefont {Saito},\ and\
  \citenamefont {Asai}}]{Lee_depth_2023}%
  \BibitemOpen
  \bibfield  {author} {\bibinfo {author} {\bibfnamefont {X.}~\bibnamefont
  {Lee}}, \bibinfo {author} {\bibfnamefont {N.}~\bibnamefont {Xie}}, \bibinfo
  {author} {\bibfnamefont {D.}~\bibnamefont {Cai}}, \bibinfo {author}
  {\bibfnamefont {Y.}~\bibnamefont {Saito}},\ and\ \bibinfo {author}
  {\bibfnamefont {N.}~\bibnamefont {Asai}},\ }\bibfield  {title} {\bibinfo
  {title} {A depth-progressive initialization strategy for quantum approximate
  optimization algorithm},\ }\bibfield  {journal} {\bibinfo  {journal}
  {Mathematics}\ }\textbf {\bibinfo {volume} {11}},\ \href
  {https://doi.org/10.3390/math11092176} {10.3390/math11092176} (\bibinfo
  {year} {2023})\BibitemShut {NoStop}%
\bibitem [{\citenamefont {Shaydulin}\ \emph {et~al.}(2023)\citenamefont
  {Shaydulin}, \citenamefont {Lotshaw}, \citenamefont {Larson}, \citenamefont
  {Ostrowski},\ and\ \citenamefont {Humble}}]{Shaydulin_parameter_2023}%
  \BibitemOpen
  \bibfield  {author} {\bibinfo {author} {\bibfnamefont {R.}~\bibnamefont
  {Shaydulin}}, \bibinfo {author} {\bibfnamefont {P.~C.}\ \bibnamefont
  {Lotshaw}}, \bibinfo {author} {\bibfnamefont {J.}~\bibnamefont {Larson}},
  \bibinfo {author} {\bibfnamefont {J.}~\bibnamefont {Ostrowski}},\ and\
  \bibinfo {author} {\bibfnamefont {T.~S.}\ \bibnamefont {Humble}},\ }\bibfield
   {title} {\bibinfo {title} {Parameter transfer for quantum approximate
  optimization of weighted {{MaxCut}}},\ }\bibfield  {journal} {\bibinfo
  {journal} {ACM Transactions on Quantum Computing}\ }\textbf {\bibinfo
  {volume} {4}},\ \href {https://doi.org/10.1145/3584706} {10.1145/3584706}
  (\bibinfo {year} {2023})\BibitemShut {NoStop}%
\bibitem [{\citenamefont {Deshpande}\ and\ \citenamefont
  {Melnikov}(2022)}]{Deshpande_capturing_2022}%
  \BibitemOpen
  \bibfield  {author} {\bibinfo {author} {\bibfnamefont {A.}~\bibnamefont
  {Deshpande}}\ and\ \bibinfo {author} {\bibfnamefont {A.}~\bibnamefont
  {Melnikov}},\ }\bibfield  {title} {\bibinfo {title} {Capturing symmetries of
  quantum optimization algorithms using graph neural networks},\ }\bibfield
  {journal} {\bibinfo  {journal} {Symmetry}\ }\textbf {\bibinfo {volume}
  {14}},\ \href {https://doi.org/10.3390/sym14122593} {10.3390/sym14122593}
  (\bibinfo {year} {2022})\BibitemShut {NoStop}%
\bibitem [{\citenamefont {Cheng}\ \emph {et~al.}(2024)\citenamefont {Cheng},
  \citenamefont {Chen}, \citenamefont {Zhang},\ and\ \citenamefont
  {Zhang}}]{cheng_quantum_2024}%
  \BibitemOpen
  \bibfield  {author} {\bibinfo {author} {\bibfnamefont {L.}~\bibnamefont
  {Cheng}}, \bibinfo {author} {\bibfnamefont {Y.-Q.}\ \bibnamefont {Chen}},
  \bibinfo {author} {\bibfnamefont {S.-X.}\ \bibnamefont {Zhang}},\ and\
  \bibinfo {author} {\bibfnamefont {S.}~\bibnamefont {Zhang}},\ }\bibfield
  {title} {\bibinfo {title} {Quantum approximate optimization via
  learning-based adaptive optimization},\ }\href
  {https://doi.org/10.1038/s42005-024-01577-x} {\bibfield  {journal} {\bibinfo
  {journal} {Communications Physics}\ }\textbf {\bibinfo {volume} {7}},\
  \bibinfo {pages} {83} (\bibinfo {year} {2024})}\BibitemShut {NoStop}%
\bibitem [{\citenamefont {Acampora}\ \emph {et~al.}(2023)\citenamefont
  {Acampora}, \citenamefont {Chiatto},\ and\ \citenamefont
  {Vitiello}}]{giovanni_genetic_2023}%
  \BibitemOpen
  \bibfield  {author} {\bibinfo {author} {\bibfnamefont {G.}~\bibnamefont
  {Acampora}}, \bibinfo {author} {\bibfnamefont {A.}~\bibnamefont {Chiatto}},\
  and\ \bibinfo {author} {\bibfnamefont {A.}~\bibnamefont {Vitiello}},\
  }\bibfield  {title} {\bibinfo {title} {Genetic algorithms as classical
  optimizer for the quantum approximate optimization algorithm},\ }\href
  {https://doi.org/10.1016/j.asoc.2023.110296} {\bibfield  {journal} {\bibinfo
  {journal} {Applied Soft Computing}\ }\textbf {\bibinfo {volume} {142}},\
  \bibinfo {pages} {110296} (\bibinfo {year} {2023})}\BibitemShut {NoStop}%
\bibitem [{\citenamefont {Bittel}\ and\ \citenamefont
  {Kliesch}(2021)}]{Bittel_training_2021}%
  \BibitemOpen
  \bibfield  {author} {\bibinfo {author} {\bibfnamefont {L.}~\bibnamefont
  {Bittel}}\ and\ \bibinfo {author} {\bibfnamefont {M.}~\bibnamefont
  {Kliesch}},\ }\bibfield  {title} {\bibinfo {title} {Training variational
  quantum algorithms is {{NP-hard}}},\ }\href
  {https://doi.org/10.1103/PhysRevLett.127.120502} {\bibfield  {journal}
  {\bibinfo  {journal} {Physical Review Letters}\ }\textbf {\bibinfo {volume}
  {127}},\ \bibinfo {pages} {120502} (\bibinfo {year} {2021})}\BibitemShut
  {NoStop}%
\bibitem [{\citenamefont {Farhi}\ \emph
  {et~al.}(2022{\natexlab{b}})\citenamefont {Farhi}, \citenamefont {Goldstone},
  \citenamefont {Gutmann},\ and\ \citenamefont {Zhou}}]{Farhi_2022}%
  \BibitemOpen
  \bibfield  {author} {\bibinfo {author} {\bibfnamefont {E.}~\bibnamefont
  {Farhi}}, \bibinfo {author} {\bibfnamefont {J.}~\bibnamefont {Goldstone}},
  \bibinfo {author} {\bibfnamefont {S.}~\bibnamefont {Gutmann}},\ and\ \bibinfo
  {author} {\bibfnamefont {L.}~\bibnamefont {Zhou}},\ }\bibfield  {title}
  {\bibinfo {title} {The quantum approximate optimization algorithm and the
  {S}herrington-{K}irkpatrick model at infinite size},\ }\href
  {https://doi.org/10.22331/q-2022-07-07-759} {\bibfield  {journal} {\bibinfo
  {journal} {Quantum}\ }\textbf {\bibinfo {volume} {6}},\ \bibinfo {pages}
  {759} (\bibinfo {year} {2022}{\natexlab{b}})}\BibitemShut {NoStop}%
\bibitem [{\citenamefont {Basso}\ \emph {et~al.}(2022)\citenamefont {Basso},
  \citenamefont {Farhi}, \citenamefont {Marwaha}, \citenamefont {Villalonga},\
  and\ \citenamefont {Zhou}}]{Basso_quantum_2022}%
  \BibitemOpen
  \bibfield  {author} {\bibinfo {author} {\bibfnamefont {J.}~\bibnamefont
  {Basso}}, \bibinfo {author} {\bibfnamefont {E.}~\bibnamefont {Farhi}},
  \bibinfo {author} {\bibfnamefont {K.}~\bibnamefont {Marwaha}}, \bibinfo
  {author} {\bibfnamefont {B.}~\bibnamefont {Villalonga}},\ and\ \bibinfo
  {author} {\bibfnamefont {L.}~\bibnamefont {Zhou}},\ }\bibfield  {title}
  {\bibinfo {title} {{The Quantum Approximate Optimization Algorithm at High
  Depth for MaxCut on Large-Girth Regular Graphs and the
  Sherrington-Kirkpatrick Model}},\ }in\ \href
  {https://doi.org/10.4230/LIPIcs.TQC.2022.7} {\emph {\bibinfo {booktitle}
  {17th Conference on the Theory of Quantum Computation, Communication and
  Cryptography (TQC 2022)}}},\ \bibinfo {series} {Leibniz International
  Proceedings in Informatics (LIPIcs)}, Vol.\ \bibinfo {volume} {232},\
  \bibinfo {editor} {edited by\ \bibinfo {editor} {\bibfnamefont
  {F.}~\bibnamefont {Le~Gall}}\ and\ \bibinfo {editor} {\bibfnamefont
  {T.}~\bibnamefont {Morimae}}}\ (\bibinfo  {publisher} {Schloss Dagstuhl --
  Leibniz-Zentrum f{\"u}r Informatik},\ \bibinfo {address} {Dagstuhl,
  Germany},\ \bibinfo {year} {2022})\ pp.\ \bibinfo {pages}
  {7:1--7:21}\BibitemShut {NoStop}%
\bibitem [{\citenamefont {Akshay}\ \emph {et~al.}(2021)\citenamefont {Akshay},
  \citenamefont {Rabinovich}, \citenamefont {Campos},\ and\ \citenamefont
  {Biamonte}}]{Akshay_parameter_2021}%
  \BibitemOpen
  \bibfield  {author} {\bibinfo {author} {\bibfnamefont {V.}~\bibnamefont
  {Akshay}}, \bibinfo {author} {\bibfnamefont {D.}~\bibnamefont {Rabinovich}},
  \bibinfo {author} {\bibfnamefont {E.}~\bibnamefont {Campos}},\ and\ \bibinfo
  {author} {\bibfnamefont {J.}~\bibnamefont {Biamonte}},\ }\bibfield  {title}
  {\bibinfo {title} {Parameter concentrations in quantum approximate
  optimization},\ }\href {https://doi.org/10.1103/PhysRevA.104.L010401}
  {\bibfield  {journal} {\bibinfo  {journal} {Physical Review A: Atomic,
  Molecular, and Optical Physics}\ }\textbf {\bibinfo {volume} {104}},\
  \bibinfo {pages} {L010401} (\bibinfo {year} {2021})}\BibitemShut {NoStop}%
\bibitem [{\citenamefont {Claes}\ and\ \citenamefont {van
  Dam}(2021)}]{Claes_2021}%
  \BibitemOpen
  \bibfield  {author} {\bibinfo {author} {\bibfnamefont {J.}~\bibnamefont
  {Claes}}\ and\ \bibinfo {author} {\bibfnamefont {W.}~\bibnamefont {van
  Dam}},\ }\bibfield  {title} {\bibinfo {title} {Instance independence of
  single layer quantum approximate optimization algorithm on mixed-spin models
  at infinite size},\ }\href {https://doi.org/10.22331/q-2021-09-15-542}
  {\bibfield  {journal} {\bibinfo  {journal} {Quantum}\ }\textbf {\bibinfo
  {volume} {5}},\ \bibinfo {pages} {542} (\bibinfo {year} {2021})}\BibitemShut
  {NoStop}%
\bibitem [{\citenamefont {{D{\'{\i}}ez-Valle}}\ \emph
  {et~al.}(2023)\citenamefont {{D{\'{\i}}ez-Valle}}, \citenamefont {Porras},\
  and\ \citenamefont {{Garc{\'{\i}}a-Ripoll}}}]{DiezValle_PRL23}%
  \BibitemOpen
  \bibfield  {author} {\bibinfo {author} {\bibfnamefont {P.}~\bibnamefont
  {{D{\'{\i}}ez-Valle}}}, \bibinfo {author} {\bibfnamefont {D.}~\bibnamefont
  {Porras}},\ and\ \bibinfo {author} {\bibfnamefont {J.~J.}\ \bibnamefont
  {{Garc{\'{\i}}a-Ripoll}}},\ }\bibfield  {title} {\bibinfo {title} {Quantum
  approximate optimization algorithm pseudo-boltzmann states},\ }\href
  {https://doi.org/10.1103/PhysRevLett.130.050601} {\bibfield  {journal}
  {\bibinfo  {journal} {Physical Review Letters}\ }\textbf {\bibinfo {volume}
  {130}},\ \bibinfo {pages} {050601} (\bibinfo {year} {2023})}\BibitemShut
  {NoStop}%
\bibitem [{\citenamefont {{D{\'i}ez-Valle}}\ \emph {et~al.}(2024)\citenamefont
  {{D{\'i}ez-Valle}}, \citenamefont {Porras},\ and\ \citenamefont
  {{Garc{\'i}a-Ripoll}}}]{DiezValle_Frontiers2024}%
  \BibitemOpen
  \bibfield  {author} {\bibinfo {author} {\bibfnamefont {P.}~\bibnamefont
  {{D{\'i}ez-Valle}}}, \bibinfo {author} {\bibfnamefont {D.}~\bibnamefont
  {Porras}},\ and\ \bibinfo {author} {\bibfnamefont {J.~J.}\ \bibnamefont
  {{Garc{\'i}a-Ripoll}}},\ }\bibfield  {title} {\bibinfo {title} {Connection
  between single-layer quantum approximate optimization algorithm
  interferometry and thermal distribution sampling},\ }\href
  {https://doi.org/10.3389/frqst.2024.1321264} {\bibfield  {journal} {\bibinfo
  {journal} {Frontiers in Quantum Science and Technology}\ }\textbf {\bibinfo
  {volume} {3}},\ \bibinfo {pages} {1321264} (\bibinfo {year}
  {2024})}\BibitemShut {NoStop}%
\bibitem [{\citenamefont {Galda}\ \emph {et~al.}(2021)\citenamefont {Galda},
  \citenamefont {Liu}, \citenamefont {Lykov}, \citenamefont {Alexeev},\ and\
  \citenamefont {Safro}}]{Galsa_transferability_2021}%
  \BibitemOpen
  \bibfield  {author} {\bibinfo {author} {\bibfnamefont {A.}~\bibnamefont
  {Galda}}, \bibinfo {author} {\bibfnamefont {X.}~\bibnamefont {Liu}}, \bibinfo
  {author} {\bibfnamefont {D.}~\bibnamefont {Lykov}}, \bibinfo {author}
  {\bibfnamefont {Y.}~\bibnamefont {Alexeev}},\ and\ \bibinfo {author}
  {\bibfnamefont {I.}~\bibnamefont {Safro}},\ }\bibfield  {title} {\bibinfo
  {title} {Transferability of optimal {{QAOA}} parameters between random
  graphs},\ }in\ \href {https://doi.org/10.1109/QCE52317.2021.00034} {\emph
  {\bibinfo {booktitle} {2021 {{IEEE}} International Conference on Quantum
  Computing and Engineering ({{QCE}})}}}\ (\bibinfo {year} {2021})\ pp.\
  \bibinfo {pages} {171--180}\BibitemShut {NoStop}%
\bibitem [{\citenamefont {Shaydulin}\ \emph {et~al.}(2024)\citenamefont
  {Shaydulin}, \citenamefont {Li}, \citenamefont {Chakrabarti}, \citenamefont
  {DeCross}, \citenamefont {Herman}, \citenamefont {Kumar}, \citenamefont
  {Larson}, \citenamefont {Lykov}, \citenamefont {Minssen}, \citenamefont
  {Sun}, \citenamefont {Alexeev}, \citenamefont {Dreiling}, \citenamefont
  {Gaebler}, \citenamefont {Gatterman}, \citenamefont {Gerber}, \citenamefont
  {Gilmore}, \citenamefont {Gresh}, \citenamefont {Hewitt}, \citenamefont
  {Horst}, \citenamefont {Hu}, \citenamefont {Johansen}, \citenamefont
  {Matheny}, \citenamefont {Mengle}, \citenamefont {Mills}, \citenamefont
  {Moses}, \citenamefont {Neyenhuis}, \citenamefont {Siegfried}, \citenamefont
  {Yalovetzky},\ and\ \citenamefont {Pistoia}}]{shaydulin2024}%
  \BibitemOpen
  \bibfield  {author} {\bibinfo {author} {\bibfnamefont {R.}~\bibnamefont
  {Shaydulin}}, \bibinfo {author} {\bibfnamefont {C.}~\bibnamefont {Li}},
  \bibinfo {author} {\bibfnamefont {S.}~\bibnamefont {Chakrabarti}}, \bibinfo
  {author} {\bibfnamefont {M.}~\bibnamefont {DeCross}}, \bibinfo {author}
  {\bibfnamefont {D.}~\bibnamefont {Herman}}, \bibinfo {author} {\bibfnamefont
  {N.}~\bibnamefont {Kumar}}, \bibinfo {author} {\bibfnamefont
  {J.}~\bibnamefont {Larson}}, \bibinfo {author} {\bibfnamefont
  {D.}~\bibnamefont {Lykov}}, \bibinfo {author} {\bibfnamefont
  {P.}~\bibnamefont {Minssen}}, \bibinfo {author} {\bibfnamefont
  {Y.}~\bibnamefont {Sun}}, \bibinfo {author} {\bibfnamefont {Y.}~\bibnamefont
  {Alexeev}}, \bibinfo {author} {\bibfnamefont {J.~M.}\ \bibnamefont
  {Dreiling}}, \bibinfo {author} {\bibfnamefont {J.~P.}\ \bibnamefont
  {Gaebler}}, \bibinfo {author} {\bibfnamefont {T.~M.}\ \bibnamefont
  {Gatterman}}, \bibinfo {author} {\bibfnamefont {J.~A.}\ \bibnamefont
  {Gerber}}, \bibinfo {author} {\bibfnamefont {K.}~\bibnamefont {Gilmore}},
  \bibinfo {author} {\bibfnamefont {D.}~\bibnamefont {Gresh}}, \bibinfo
  {author} {\bibfnamefont {N.}~\bibnamefont {Hewitt}}, \bibinfo {author}
  {\bibfnamefont {C.~V.}\ \bibnamefont {Horst}}, \bibinfo {author}
  {\bibfnamefont {S.}~\bibnamefont {Hu}}, \bibinfo {author} {\bibfnamefont
  {J.}~\bibnamefont {Johansen}}, \bibinfo {author} {\bibfnamefont
  {M.}~\bibnamefont {Matheny}}, \bibinfo {author} {\bibfnamefont
  {T.}~\bibnamefont {Mengle}}, \bibinfo {author} {\bibfnamefont
  {M.}~\bibnamefont {Mills}}, \bibinfo {author} {\bibfnamefont {S.~A.}\
  \bibnamefont {Moses}}, \bibinfo {author} {\bibfnamefont {B.}~\bibnamefont
  {Neyenhuis}}, \bibinfo {author} {\bibfnamefont {P.}~\bibnamefont
  {Siegfried}}, \bibinfo {author} {\bibfnamefont {R.}~\bibnamefont
  {Yalovetzky}},\ and\ \bibinfo {author} {\bibfnamefont {M.}~\bibnamefont
  {Pistoia}},\ }\bibfield  {title} {\bibinfo {title} {Evidence of scaling
  advantage for the quantum approximate optimization algorithm on a classically
  intractable problem},\ }\href {https://doi.org/10.1126/sciadv.adm6761}
  {\bibfield  {journal} {\bibinfo  {journal} {Science Advances}\ }\textbf
  {\bibinfo {volume} {10}},\ \bibinfo {pages} {eadm6761} (\bibinfo {year}
  {2024})}\BibitemShut {NoStop}%
\bibitem [{\citenamefont {Sureshbabu}\ \emph {et~al.}(2024)\citenamefont
  {Sureshbabu}, \citenamefont {Herman}, \citenamefont {Shaydulin},
  \citenamefont {Basso}, \citenamefont {Chakrabarti}, \citenamefont {Sun},\
  and\ \citenamefont {Pistoia}}]{Sureshbabu_2024}%
  \BibitemOpen
  \bibfield  {author} {\bibinfo {author} {\bibfnamefont {S.~H.}\ \bibnamefont
  {Sureshbabu}}, \bibinfo {author} {\bibfnamefont {D.}~\bibnamefont {Herman}},
  \bibinfo {author} {\bibfnamefont {R.}~\bibnamefont {Shaydulin}}, \bibinfo
  {author} {\bibfnamefont {J.}~\bibnamefont {Basso}}, \bibinfo {author}
  {\bibfnamefont {S.}~\bibnamefont {Chakrabarti}}, \bibinfo {author}
  {\bibfnamefont {Y.}~\bibnamefont {Sun}},\ and\ \bibinfo {author}
  {\bibfnamefont {M.}~\bibnamefont {Pistoia}},\ }\bibfield  {title} {\bibinfo
  {title} {Parameter setting in quantum approximate optimization of weighted
  problems},\ }\href {https://doi.org/10.22331/q-2024-01-18-1231} {\bibfield
  {journal} {\bibinfo  {journal} {Quantum}\ }\textbf {\bibinfo {volume} {8}},\
  \bibinfo {pages} {1231} (\bibinfo {year} {2024})}\BibitemShut {NoStop}%
\bibitem [{\citenamefont {Kochenberger}\ \emph {et~al.}(2014)\citenamefont
  {Kochenberger}, \citenamefont {Hao}, \citenamefont {Glover}, \citenamefont
  {Lewis}, \citenamefont {L{\"u}}, \citenamefont {Wang},\ and\ \citenamefont
  {Wang}}]{kochenberger2014}%
  \BibitemOpen
  \bibfield  {author} {\bibinfo {author} {\bibfnamefont {G.}~\bibnamefont
  {Kochenberger}}, \bibinfo {author} {\bibfnamefont {J.-K.}\ \bibnamefont
  {Hao}}, \bibinfo {author} {\bibfnamefont {F.}~\bibnamefont {Glover}},
  \bibinfo {author} {\bibfnamefont {M.}~\bibnamefont {Lewis}}, \bibinfo
  {author} {\bibfnamefont {Z.}~\bibnamefont {L{\"u}}}, \bibinfo {author}
  {\bibfnamefont {H.}~\bibnamefont {Wang}},\ and\ \bibinfo {author}
  {\bibfnamefont {Y.}~\bibnamefont {Wang}},\ }\bibfield  {title} {\bibinfo
  {title} {The unconstrained binary quadratic programming problem: A survey},\
  }\href {https://doi.org/10.1007/s10878-014-9734-0} {\bibfield  {journal}
  {\bibinfo  {journal} {Journal of Combinatorial Optimization}\ }\textbf
  {\bibinfo {volume} {28}},\ \bibinfo {pages} {58} (\bibinfo {year}
  {2014})}\BibitemShut {NoStop}%
\bibitem [{\citenamefont {Lidar}(2008)}]{Lidar_PRL08}%
  \BibitemOpen
  \bibfield  {author} {\bibinfo {author} {\bibfnamefont {D.~A.}\ \bibnamefont
  {Lidar}},\ }\bibfield  {title} {\bibinfo {title} {Towards fault tolerant
  adiabatic quantum computation},\ }\href
  {https://doi.org/10.1103/PhysRevLett.100.160506} {\bibfield  {journal}
  {\bibinfo  {journal} {Physical Review Letters}\ }\textbf {\bibinfo {volume}
  {100}},\ \bibinfo {pages} {160506} (\bibinfo {year} {2008})}\BibitemShut
  {NoStop}%
\bibitem [{\citenamefont {Wu}\ \emph {et~al.}(2002)\citenamefont {Wu},
  \citenamefont {Byrd},\ and\ \citenamefont {Lidar}}]{Lidar_PRL02}%
  \BibitemOpen
  \bibfield  {author} {\bibinfo {author} {\bibfnamefont {L.-A.}\ \bibnamefont
  {Wu}}, \bibinfo {author} {\bibfnamefont {M.~S.}\ \bibnamefont {Byrd}},\ and\
  \bibinfo {author} {\bibfnamefont {D.~A.}\ \bibnamefont {Lidar}},\ }\bibfield
  {title} {\bibinfo {title} {Polynomial-time simulation of pairing models on a
  quantum computer},\ }\href {https://doi.org/10.1103/PhysRevLett.89.057904}
  {\bibfield  {journal} {\bibinfo  {journal} {Physical Review Letters}\
  }\textbf {\bibinfo {volume} {89}},\ \bibinfo {pages} {057904} (\bibinfo
  {year} {2002})}\BibitemShut {NoStop}%
\bibitem [{\citenamefont {Lotshaw}\ \emph {et~al.}(2023)\citenamefont
  {Lotshaw}, \citenamefont {Siopsis}, \citenamefont {Ostrowski}, \citenamefont
  {Herrman}, \citenamefont {Alam}, \citenamefont {Powers},\ and\ \citenamefont
  {Humble}}]{Lotshaw_PRA23}%
  \BibitemOpen
  \bibfield  {author} {\bibinfo {author} {\bibfnamefont {P.~C.}\ \bibnamefont
  {Lotshaw}}, \bibinfo {author} {\bibfnamefont {G.}~\bibnamefont {Siopsis}},
  \bibinfo {author} {\bibfnamefont {J.}~\bibnamefont {Ostrowski}}, \bibinfo
  {author} {\bibfnamefont {R.}~\bibnamefont {Herrman}}, \bibinfo {author}
  {\bibfnamefont {R.}~\bibnamefont {Alam}}, \bibinfo {author} {\bibfnamefont
  {S.}~\bibnamefont {Powers}},\ and\ \bibinfo {author} {\bibfnamefont {T.~S.}\
  \bibnamefont {Humble}},\ }\bibfield  {title} {\bibinfo {title} {Approximate
  {{Boltzmann}} distributions in quantum approximate optimization},\ }\href
  {https://doi.org/10.1103/PhysRevA.108.042411} {\bibfield  {journal} {\bibinfo
   {journal} {Physical Review A: Atomic, Molecular, and Optical Physics}\
  }\textbf {\bibinfo {volume} {108}},\ \bibinfo {pages} {042411} (\bibinfo
  {year} {2023})}\BibitemShut {NoStop}%
\bibitem [{\citenamefont {Boulebnane}\ \emph {et~al.}(2025)\citenamefont
  {Boulebnane}, \citenamefont {Sud}, \citenamefont {Shaydulin},\ and\
  \citenamefont {Pistoia}}]{boulebnane2025}%
  \BibitemOpen
  \bibfield  {author} {\bibinfo {author} {\bibfnamefont {S.}~\bibnamefont
  {Boulebnane}}, \bibinfo {author} {\bibfnamefont {J.}~\bibnamefont {Sud}},
  \bibinfo {author} {\bibfnamefont {R.}~\bibnamefont {Shaydulin}},\ and\
  \bibinfo {author} {\bibfnamefont {M.}~\bibnamefont {Pistoia}},\ }\href
  {https://arxiv.org/abs/2503.09563} {\bibinfo {title} {Equivalence of quantum
  approximate optimization algorithm and linear-time quantum annealing for the
  {S}herrington-{K}irkpatrick model}} (\bibinfo {year} {2025}),\ \Eprint
  {https://arxiv.org/abs/2503.09563} {arXiv:2503.09563 [quant-ph]} \BibitemShut
  {NoStop}%
\bibitem [{\citenamefont {Chen}\ and\ \citenamefont
  {Zhang}(2024)}]{chen_effective_2024}%
  \BibitemOpen
  \bibfield  {author} {\bibinfo {author} {\bibfnamefont {Y.-Q.}\ \bibnamefont
  {Chen}}\ and\ \bibinfo {author} {\bibfnamefont {S.-X.}\ \bibnamefont
  {Zhang}},\ }\href {https://doi.org/10.48550/ARXIV.2411.18921} {\bibinfo
  {title} {Effective temperature in approximate quantum many-body states}}
  (\bibinfo {year} {2024}),\ \Eprint {https://arxiv.org/abs/2411.18921}
  {arXiv:2411.18921 [quant-ph]} \BibitemShut {NoStop}%
\bibitem [{\citenamefont {Virtanen~{\it et al}}(2020)}]{2020SciPy-NMeth}%
  \BibitemOpen
  \bibfield  {author} {\bibinfo {author} {\bibfnamefont {P.}~\bibnamefont
  {Virtanen~{\it et al}}},\ }\bibfield  {title} {\bibinfo {title} {{{SciPy}
  1.0: Fundamental Algorithms for Scientific Computing in Python}},\ }\href
  {https://doi.org/10.1038/s41592-019-0686-2} {\bibfield  {journal} {\bibinfo
  {journal} {Nature Methods}\ }\textbf {\bibinfo {volume} {17}},\ \bibinfo
  {pages} {261} (\bibinfo {year} {2020})}\BibitemShut {NoStop}%
\end{thebibliography}%
\end{document}